\documentclass[fleqn]{aa}
\usepackage{graphicx}
\usepackage{amssymb}
\usepackage{amsmath}
\usepackage{natbib}
\usepackage{mathrsfs}
\usepackage{ulem}
\usepackage{txfonts}
\usepackage{lmodern}
\usepackage{booktabs}
\usepackage{multicol}
\usepackage{cases}
\usepackage{color}

\usepackage[switch]{lineno} 

\newcommand{\corot}{{\textsc{CoRoT}}}

\newcommand{\kepler}{\textit{Kepler}}

\def\m2s2{\,m$^{2}$\,s$^{-2}$} 

\newcommand{\vect}[1]{\boldsymbol{\rm #1}}

\newcommand{\vaisala}{Brunt-V\"ais\"al\"a}

\newcommand{\mesa}{\textsc{MESA}}

\newcommand{\adipls}{\textsc{ADIPLS}}
\newcommand{\gyre}{\textsc{GYRE}}

\newcommand{\deriv}[2]{\frac{\hbox{d} #1}{\hbox{d} #2}}
\newcommand{\dderiv}[2]{\frac{\hbox{d}^2 #1}{\hbox{d} #2 ^2}}

\newcommand{\ra}{r_{\rm a}}
\newcommand{\rb}{r_{\rm b}}
\newcommand{\rc}{r_{\rm c}}
\newcommand{\rd}{r_{\rm d}}
\newcommand{\re}{r_{\rm e}}
\newcommand{\rf}{r_{\rm f}}
\newcommand{\gun}{g$_1$}
\newcommand{\gdeux}{g$_2$}
\newcommand{\dr}{\hbox{d}r}
\newcommand{\tgun}{\theta_{\rm g_1}}
\newcommand{\tgdeux}{\theta_{\rm g_2}}
\newcommand{\tetap}{\theta_{\rm p}}
\newcommand{\cs}{c_{\rm s}}

\newenvironment{itemize*}%
  {\begin{itemize}%
    \setlength{\itemsep}{1pt}%
    \setlength{\parskip}{1pt}}%
  {\end{itemize}}

\newcommand\T{\rule{0pt}{2.6ex}}
\newcommand\B{\rule[-1.2ex]{0pt}{0pt}}

\begin{document}
\title{Seismic characterization of red giants going through the Helium-core flash}
\titlerunning{Seismic characterization of red giants going through the Helium-core flash}
\author{
S. Deheuvels\inst{1}
\and K. Belkacem\inst{2}
}

\institute{IRAP, Universit\'e de Toulouse, CNRS, CNES, UPS, (Toulouse), France
\and LESIA, Observatoire de Paris, PSL Research University, CNRS, Universit\'e Pierre et Marie Curie, Universit\'e Denis Diderot, 92195 Meudon, France}

\offprints{S. Deheuvels\\ \email{sebastien.deheuvels@irap.omp.eu}
}


\abstract{First-ascent red giants in the approximate mass range $0.7\lesssim M/M_\odot\lesssim 2$ ignite helium in their degenerate core as a flash. Stellar evolution codes predict that the He flash consists of a series of consecutive subflashes. Observational evidence of the existence of the He flash and subflashes is lacking. The detection of mixed modes in red giants from space missions \corot\ and \kepler\ has opened new opportunities to search for such evidence.}
{During a subflash, the He burning shell is convective, which splits the cavity of gravity modes in two. We here investigate how this additional cavity modifies the oscillation spectrum of the star. We also address the question of the detectability of the modes, to determine whether they could be used to seismically identify red giants passing through the He flash.}
{We calculate the asymptotic mode frequencies of stellar models going through a He subflash using the JWKB approximation. To predict the detectability of the modes, we estimate their expected heights, taking into account the effects of radiative damping in the core. Our results are then compared to the oscillation spectra obtained by calculating numerically the mode frequencies during a He subflash.}
{We show that during a He subflash, the detectable oscillation spectrum mainly consists of modes trapped in the acoustic cavity and in the outer g-mode cavity. The spectrum should thus at first sight resemble that of a core-helium-burning giant. However, we find a list of clear, detectable features that could enable us to identify red giants passing through a He subflash. In particular, during a He subflash, several modes that are trapped in the innermost g-mode cavity are expected to be detectable. We show that these modes could be identified by their frequencies or by their rotational splittings. Other features, such as the measured period spacing of gravity modes or the location of the H-burning shell within the g-mode cavity could also be used to identify stars going through a He subflash.}
%
{The features derived in this study can now be searched for in the large datasets provided by the \corot\ and \kepler\ missions.}


\maketitle

\section{Introduction \label{sect_intro}}

Stars in the mass range $0.7\lesssim M/M_\odot\lesssim 2$ ignite He in their core under conditions of strong electron degeneracy, which results in a thermal runaway known as the He core flash. 
The basic features of the He core flash have been known since numerical models of stellar evolution were followed from the tip of the red giant branch to the core He burning phase (\citealt{harm64}, \citealt{thomas67}). It is well established that He is ignited off-center, in the layers of maximal temperature, because of the energy carried away by neutrinos in the center of the star. Instead of expanding and cooling, these layers are further heated owing to the degeneracy of electrons, which results in a thermal runaway. The localized heating leads to superadiabatic gradients and to the onset of convection in the He-burning layers. The rapid increase in temperature eventually removes the electron degeneracy. 
Initial hydrodynamic calculations of the He core flash found that it should induce a disruption of the star (\citealt{edwards69}, \citealt{deupree84}), while more modern 2D- and 3D-simulations have shown that the flash does not produce a hydrodynamical event (\citealt{deupree96}, \citealt{mocak08}, \citealt{mocak09}). 

One critical question is the time it takes for He burning to reach the center of the star. 1D evolutionary models predict that after the peak of the He flash, electron degeneracy is locally removed in the shell where He was ignited. The layers below remain inert and degenerate until the heat produced by the first He flash diffuses inward. Electron degeneracy is then removed in the inner layers through a series of weaker secondary He flashes occurring closer and closer to the stellar center (\citealt{thomas67}, \citealt{iben84}, \citealt{bildsten12}). The duration of the phase of successive He subflashes is determined by the timescale over which thermal diffusion operates inward after each subflash. It was found to be of the order of 2 Myr (\citealt{bildsten12}), which represents a non-negligible fraction of the duration of the phase of quiet He core burning ($\sim$ 100 Myr). The existence of these subflashes has been questioned by 2D- and 3D-simulations of the He core flash (\citealt{mocak08}, \citealt{mocak09}). These studies suggest that the convective region that develops as a result of He burning rapidly extends inward, potentially reaching the center of the star on a timescale of about a month. Degeneracy would then be lifted in the core without the occurrence of He subflashes. So far, no observational evidence was obtained of any star going through the He-flash or the subsequent subflashes. 


If the He flash consists in a series of subflashes, the odds of observing a star in this phase are much higher than if it is a single event, as suggested by 2D- and 3D-simulations. Asteroseismology could then be very helpful to identify stars during the He flash. Indeed, red giants are known to stochastically excite non-radial mixed modes in their convective envelopes. These modes behave as gravity (g) modes in the core, and as pressure (p) modes in the envelope. The exceptional diagnostic potential of these modes has been known since they were found in stellar models (\citealt{dziembowski71}, \citealt{scuflaire74}, \citealt{osaki75}). With the advent of space missions \corot\ (\citealt{baglin06}) and \kepler\ (\citealt{borucki10}), mixed modes were detected in thousands of subgiants (\citealt{deheuvels10}, \citealt{deheuvels11}, \citealt{campante11}) and red giants (\citealt{bedding11}, \citealt{mosser11}). Among other applications, mixed modes can be used to measure the nearly constant period spacing $\Delta\Pi$ of high-radial-order dipolar g modes (\citealt{bedding11}, \citealt{mosser11}, \citealt{vrard16}), which depends on the fine structure of the deep core. It has been shown that He core burning giants (belonging to the so-called red clump), which have convective cores, have distinctly larger $\Delta\Pi$ than H-shell burning giants (first-ascent red giants), whose cores are radiative. This has been used as a powerful tool to distinguish the two populations (\citealt{bedding11}, \citealt{mosser11}). \cite{bildsten12} argued that the period spacings of g modes could also be used to identify red giants in the phases between He subflashes. Indeed, in the aftermath of a He subflash, the core structure is close to that of a star on the red giant branch (RGB), but the subflash leaves an imprint that significantly modifies the period spacing of g modes. \cite{bildsten12} thus found that stars between two subflashes have values of $\Delta\Pi$ intermediate between those of RGB stars and those of He core burning giants.

In this paper, we investigate the oscillation spectrum of red giants \textit{during} the He subflashes. As mentioned above, during a subflash, the layers in which He is ignited become convective, owing to the heating that the nuclear reactions produce. Consequently, gravity waves become evanescent in the He-burning layers. The g-mode cavity is thus split into two distinct cavities that are separated by the He-burning shell. During a subflash, the star has three different cavities: two internal g-mode cavities and the external p-mode cavity. The oscillation spectrum is thus expected to be altered compared to other phases (i) because the shape of the cavities is modified, and (ii) because of the additional g-mode cavity. We here investigate whether this can be used to identify red giants undergoing a He subflash. Detecting a star in this phase would provide direct evidence of the existence of He subflashes and give us valuable insight on this poorly known event of stellar evolution.

In Sect. \ref{sect_1Dmodel}, we show that if He subflashes exist, several tens of red giants are expected to be in the process of a subflash among the $\sim$ 15,000 red giants for which the \kepler\ satellite (\citealt{borucki10}) has detected oscillations. The special case of three mode cavities inside a star has not been addressed so far. One can expect that it leads to complex oscillation spectra, whose interpretation can be complicated. In Sect. \ref{sect_JWKB}, we calculate the asymptotic mode frequencies in the case of three cavities using the JWKB approximation. We use these analytic calculations to predict the oscillation spectrum of red giants undergoing a He subflash. In Sect. \ref{sect_numerical}, we calculate numerically the oscillation spectra of stellar models during a He subflash and we compare the results to the predictions of the asymptotic mode frequencies. This leads us to propose ways of seismically identify red giants going through a He subflash in Sect. \ref{sect_discussion}.

\section{He subflashes in 1D stellar evolution models \label{sect_1Dmodel}}

\subsection{Computation of stellar models during He subflashes}

We used the stellar evolution code  \mesa\ (\citealt{paxton11}, version 10108) to evolve models of different masses ranging from 0.7 to 2 $M_\odot$ until the triggering of He-burning. We considered initial mass fractions of hydrogen of $X=0.70$ and helium $Y=0.28$, which corresponds to $Z=0.02$, and assumed the mixture of heavy elements of \cite{grevesse98}. We used the OPAL equation of state (\citealt{rogers02}), complemented by the HELM equation of state (\citealt{timmes00}) where necessary, as described by \cite{paxton11}. The opacity tables were taken from OPAL (\citealt{iglesias96}), complemented by those of \cite{ferguson05} at low temperatures. We used the nuclear reaction rates from the NACRE compilation (\citealt{angulo99}). We neglected the effects of rotation and mass loss. Convection was treated using the classical mixing-length-theory with $\alpha_{\rm MLT} = 2.0$. 

During He subflashes, sharp gradients of the mean molecular weight develop near the edges of the convective shell where He is burnt. Consequently, the extent and the evolution of the convective region depends on the criterion chosen for the onset of convection. We computed models using three different prescriptions: (i) with the Schwarzschild criterion, (ii) with the Ledoux criterion, including the effects of thermohaline mixing following the prescription of \cite{brown13}\footnote{Assuming the bare Ledoux criterion leads to numerical instabilities because of the negative gradient of mean molecular weight at the edges of the shell.}, and (iii) using the Schwarzschild criterion with a mild exponential diffusive overshooting following the prescription of \cite{herwig00}, extending over a distance of 0.01 pressure scale height. It is important to note that at the beginning of the subflash, convection starts in a region where the mean molecular weight is increasing outwards. Indeed, below the convective shell the core is comprised of nearly pure helium and above this shell, the matter has been enriched in carbon by previous subflashes. At both edges of the convective shell, the gradient of chemical composition, defined as $\nabla_\mu\equiv\hbox{d}\ln\mu/\hbox{d}\ln P$, is thus negative during the first part of the subflash, which is known to have a destabilizing effect. This is the reason why we chose to consider models that include the  \cite{brown13} prescription for thermohaline mixing, keeping in mind that the efficiency of this mixing remains uncertain. Unless mentioned otherwise, the results presented use the bare Schwarzschild criterion.

\subsection{Expected prevalence of helium-flashing giants among \kepler\ data}

\begin{figure}
\begin{center}
\includegraphics[width=9cm]{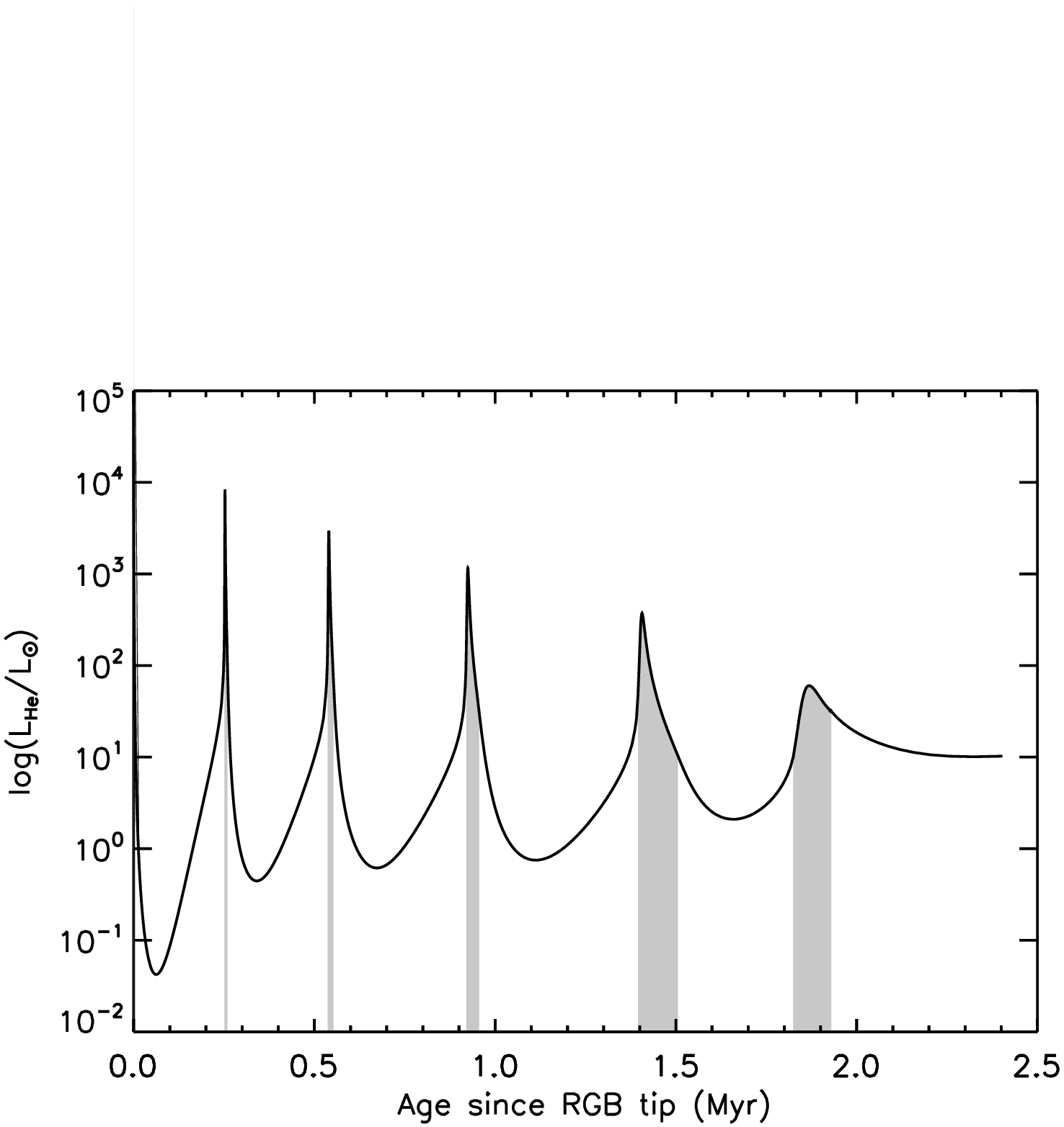}
\end{center}
\caption{Variations in the luminosity produced by He-burning during the He-subflashes of a 1.7-$M_\odot$ model computed with \mesa. The shaded areas indicate periods during which the (convective) He-burning region splits the g-mode cavity in two.
\label{fig_subflash_lum}}
\end{figure}

Before seismically characterizing red giants undergoing a He-subflash, we addressed the question of the expected number of \kepler\ targets that are in this short-lived period of their evolution. Fig. \ref{fig_subflash_lum} shows the variations in the luminosity produced by helium-burning as a function of time for a 1.7 $M_\odot$ model.
The shaded areas indicate the periods during which He-burning off-center causes the apparition of a intermediate convective region, which splits the g-mode cavity in two. The cumulative time of these subflashes corresponds to about 0.3 Myr. This represents roughly 0.4\% of the total duration of the He core burning phase (which lasts about 80 Myr). This ratio is weakly dependent on the stellar mass. This number shows that only a small fraction of primary clump red giants are in this stage of evolution. However, the \kepler\ data have led to the detection of oscillations in about 15,000 red giants. 

To estimate the expected number of helium-flashing giants, we used the catalog of \cite{vrard16}, who extracted global seismic parameters for 6,111 red giants observed with the \kepler\ satellite. About 67\% of these targets are in the He core burning phase. Among this group, about 89\% have stellar masses below $2\,M_{\odot}$, meaning that they have gone through the He flash and thus belong to the so-called \textit{primary clump}. We thus estimate to about 60\% the proportion of \kepler\ red giants that are in the primary clump. Among those, a proportion of 0.4\% is expected to be going through a subflash. This corresponds to about 35 targets among the whole \kepler\ data set. This demonstrates that, provided the He-flash occurs as a series of subflashes as predicted by 1D stellar evolution models, the \kepler\ sample should contain several tens of giants that are in the process of a subflash.
 
\subsection{Preliminary considerations on the oscillation spectra of He-flashing giants}

\begin{figure}
\begin{center}
\includegraphics[width=9cm]{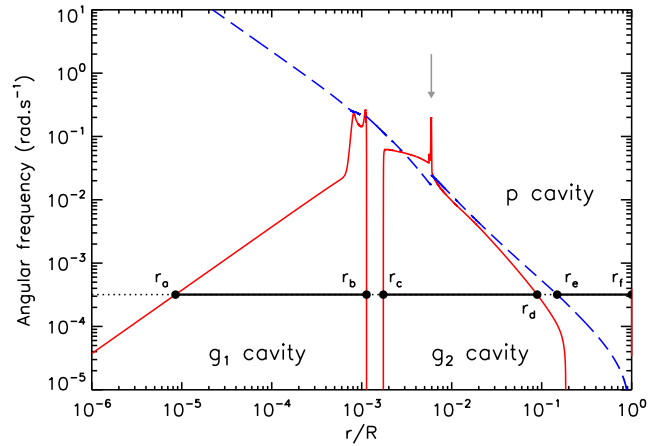}
\end{center}
\caption{Propagation diagram for a 1.7-$M_\odot$ model during a He-burning subflash. The \vaisala\ frequency is represented by the full red curve, and the $l=1$ Lamb frequency $S_1$ corresponds to the long-dashed blue curve. The horizontal line corresponds to the angular frequency $\omega=2\pi\nu_{\rm max}$. A wave with pulsation $\omega$ propagates where the line is solid, and it is evanescent where the line is dotted. Turning points for the cavities are also depicted by the filled circles. The gray vertical arrow indicates the position of the H-burning shell.
\label{fig_diag_prop}}
\end{figure}

A propagation diagram of the 1.7-$M_\odot$ model during a helium subflash is shown in Fig. \ref{fig_diag_prop}. The regions of propagation of a wave with a frequency corresponding to the expected maximum power of excited oscillation ($\nu_{\rm max}$) is overplotted. As mentioned in Sect. \ref{sect_intro}, the g-mode cavity is split in two propagating regions. We further refer to the deeper (resp. shallower) g-mode cavity as the g$_1$ (resp. g$_2$) cavity. We denote as $\ra$ and $\rb$ the inner and outer turning points of the g$_1$ cavity. Similarly, $\rc$ and $\rd$ are the turning points of the g$_2$ cavity, and $\re$ and $\rf$ are the turning points of the p-mode cavity.

During a helium subflash, the oscillation spectrum is expected to be more complex than for regular red giants, since it is a collection of the spectra of the three cavities. We also expect the frequencies of the eigenmodes of each cavity to be modified because of the coupling produced by the evanescent zones $[\rb,\rc]$ and $[\rd,\re]$. The strength of the coupling generated by the latter evanescent region, which separates the \gdeux\ cavity from the p-mode cavity, is expected to be similar to the coupling between the two cavities of regular red giants (see \citealt{mosser17}). We thus expect to detect most of the modes that are trapped mainly in the \gdeux\ cavity. The intensity of the coupling between the two g-mode cavities during a helium subflash is much more uncertain. It is also critical because if this coupling is too weak, then the g$_2$ and p-mode cavities are essentially disconnected from the inner \gun\ cavity, and we can expect to detect only mixed modes from the g$_2$ and p cavities. This would thus yield oscillation spectra qualitatively similar to those of regular red giants.

The numerical computation of mode frequencies in the red-giant phase is notoriously complex. Efforts are currently under way to compare the results of different oscillation codes for first-ascent red giants (Silva Aguirre et al. in prep.) and they are planned to be extended to clump stars. The main issues are (i) the small meshing that is needed to adequately sample the wavelengths of high-order gravity modes in the core (this usually requires to perform a re-interpolation of equilibrium quantities before calculating oscillation frequencies), and (ii) the sharp variations in the equilibrium quantities that can arise owing to the many structural changes. One can expect these issues to be magnified during the complex phase of helium subflashes. For this reason, we chose to first study the general properties of oscillations during subflashes analytically by using an asymptotic method, before attempting to numerically compute the mode frequencies in these stars. We also note that in regular red giants, asymptotic expressions of mode frequencies were extremely useful to identify the observed mixed modes and interpret the oscillation spectra (e.g. \citealt{mosser12a}). If red giants undergoing a helium subflash are ever discovered, the asymptotic expressions that are developed below in the case of three propagating cavities will be crucial to identify the observed modes.

\section{Asymptotic mode frequencies during a subflash \label{sect_JWKB}}

In this section, we extend the asymptotic method that \cite{shibahashi79} developed using the JWKB (Jeffreys-Wentzel-Kramers-Brillouin) approximation to the case of three cavities, as occurs during He-burning subflashes. 

\subsection{JWKB approximation with three propagation cavities}

We start by briefly recalling how the JWKB approximation can be applied to non-radial adiabatic oscillations, following the development proposed by \cite{shibahashi79} (see also \citealt{unno89}).
We introduce the variables $v$ and $w$ defined as
\begin{align}
& v \equiv \rho^{1/2}\cs r \left| 1-\frac{S_l^2}{\omega^2} \right| ^{-1/2} \xi_r \label{def_v} \\
& w \equiv \rho^{-1/2} r \left| N^2-\omega^2 \right| ^{-1/2} p', \label{def_w}
\end{align}
where $\xi_r$ is the radial component of the mode displacement, $p'$ is the Eulerian pressure perturbation, $\cs$ is the sound speed in the medium, $N$ is the \vaisala\ frequency, and the Lamb frequency for modes of degree $l$ is expressed as $S_l = \sqrt{l(l+1)}\cs/r$. Under the Cowling approximation, which consists in neglecting the perturbation to the gravitational potential (\citealt{cowling41}), the radial part of the equations of adiabatic stellar oscillations can be reduced to the following pair of turning-point equations (\citealt{shibahashi79}):
\begin{align}
& \dderiv{v}{r} + k_r^2 v = 0 \label{eq_v} \\
& \dderiv{w}{r} + k_r^2 w = 0, \label{eq_w}
\end{align}
where $k_r$ depends on the angular frequency $\omega$ of the wave and can be approximated as
\begin{equation}
k_r^2 \approx \frac{\omega^2}{\cs^2} \left( \frac{S_l^2}{\omega^2}-1\right) \left( \frac{N^2}{\omega^2}-1\right).
\label{eq_kr}
\end{equation}
This approximate expression neglects terms that involve derivatives of structural quantities, which are negligibly small compared to the terms in the right-hand-side of Eq. \ref{eq_kr} everywhere except near the turning points and the stellar surface.
Note also that Eq. \ref{eq_v} and \ref{eq_w} were obtained by neglecting terms involving spatial derivatives of equilibrium quantities, which are negligibly small compared to $k_r^2$ everywhere, except in the neighborhood of the turning points and near the surface. This approximation is especially valid for modes with short wavelength in the radial direction, which is the case for the modes that we are interested in.

Eq. \ref{eq_v} and \ref{eq_w} can then be solved using the JWKB approximation, i.e. assuming that the local wavelength of the wave is small compared to the scale height of variations of the medium. Using the approximate expression for $k_r$ given by Eq. \ref{eq_kr}, it is clear that turning points occur either where $\omega^2 = N^2$ or where $\omega^2 = S_l^2$. The interior of the star is thus divided in regions surrounding each turning point, and the turning-point equations are solved in each of these regions. For the turning points where $\omega^2 = N^2$, Eq. \ref{eq_w} is singular (as shown by Eq. \ref{def_w}), and we therefore solve Eq. \ref{eq_v}. Conversely, for the turning points where $\omega^2 = S_l^2$, Eq. \ref{eq_v} is singular, and we solve Eq. \ref{eq_w}. The solutions of the turning-point equations can be expressed in terms of Airy functions. The eigenfunctions of each region are then matched together, using the asymptotic forms of Airy functions. This provides a quantization condition for the eigenmodes.

Before applying the procedure described above to the case of three propagation cavities, we recall the results obtained by \cite{shibahashi79} in the cases of single cavities and double cavities.

\subsubsection{Case of a single cavity}

In the case of a single g-mode cavity, for instance assuming that the waves propagate only between $\ra$ and $\rb$ in Fig. \ref{fig_diag_prop}, the eigenfunctions must decay exponentially for $r\ll\ra$ and for $r\gg\rb$. Eq. \ref{eq_v} is solved around the turning points $\ra$ and $\rb$. 
The matching of the eigenfunctions $v$ and $w$ at an intermediate radius $r$ such that $\ra\ll r\ll \rb$ requires that
\begin{equation}
\cos\left(\int_{\ra}^{\rb} k_r\, \hbox{d}r\right) = 0,
\label{eq_gpure}
\end{equation}
thus yielding $\int_{\ra}^{\rb} k_r\, \hbox{d}r = (n+1/2)\pi$, where $n$ corresponds to the radial order. This expression can then be used to obtain approximate expressions for the frequencies of g modes (\citealt{tassoul80}).

To obtain an analogous expression for a single p-mode cavity, we assume that the wave propagates only between $\re$ and $\rf$ in Fig. \ref{fig_diag_prop}. This time, Eq. \ref{eq_w} is used around the turning point $\re$ and Eq. \ref{eq_v} around $\rf$. The matching of eigenfunctions inside the cavity provides the condtion
\begin{equation}
\sin\left(\int_{\re}^{\rf} k_r\, \hbox{d}r\right) = 0,
\label{eq_ppure}
\end{equation}
i.e. $\int_{\re}^{\rf} k_r\, \hbox{d}r = m\pi$, where $m$ corresponds to the radial order. This expression is the basis for asymptotic expressions of the frequencies of p modes (\citealt{tassoul80}).

\subsubsection{Case of two cavities}

The case of a g-mode cavity and a p-mode cavity coupled through an evanescent region has been considered by \cite{shibahashi79} and later studied extensively owing to the detection of mixed modes in red giants (e.g. \citealt{mosser12a}, \citealt{goupil13}, \citealt{jiang14}, \citealt{deheuvels15}). If we assume that the waves propagate only in the regions $[\rc,\rd]$ and $[\re,\rf]$ in Fig. \ref{fig_diag_prop} (i.e. disregarding the additional most internal g-mode cavity), one has to solve the turning-point equations in both cavities and then to match the eigenfunctions in the evanescent region in between. This matching yields the condition
\begin{equation}
\cot\left(\int_{\rc}^{\rd} k_r\, \hbox{d}r\right) \tan\left(\int_{\re}^{\rf} k_r\, \hbox{d}r\right) = \frac{1}{4} \exp \left( -2 \int_{\rd}^{\re} \kappa \dr \right)
\label{eq_g2p}
\end{equation}
where $\kappa^2=-k_r^2$ in the evanescent region. The right-hand-side term corresponds to the coupling strength between the two cavities. If this terms vanishes, we recover the conditions found for pure g and p modes given in Eq. \ref{eq_gpure} and \ref{eq_ppure}. The use of this expression was paramount to decipher the oscillation spectra of red giants (\citealt{mosser12a}, \citealt{goupil13}). We note that Eq. \ref{eq_g2p} is valid only in the case of a weak coupling between the two cavities. \cite{takata16} has derived a more general expression which extends to the case of a strong coupling and accounts for the perturbation of the gravitational potential.

\subsubsection{Case of three cavities}

We extended the derivations of \cite{shibahashi79} to the case of three cavities, as shown in Fig. \ref{fig_diag_prop}. The turning-points equations needed to be solved separately in each of the three cavities, and then matched in the two evanescent regions. Details of the calculation are given in Appendix \ref{app_JWKB}. We eventually obtained the following matching condition
\begin{equation}
\cot\tgun \cot\tgdeux \tan\tetap - q_2\cot\tgun - q_1\tan\tetap - q_1 q_2 \cot\tgdeux = 0, \label{eq_3cav}
\end{equation}
where we have defined
\begin{equation}
\tgun \equiv \int_{\ra}^{\rb} k_r \,\hbox{d}r  \;\; ; \;\; \tgdeux \equiv \int_{\rc}^{\rd} k_r \,\hbox{d}r \;\; ; \;\; \tetap \equiv \int_{\re}^{\rf} k_r \,\hbox{d}r
\end{equation}
and
\begin{align}
q_1 & \equiv \frac{1}{4} \exp \left( -2 \int_{\rb}^{\rc} \kappa \dr \right), \label{eq_q1} \\
q_2 & \equiv \frac{1}{4} \exp \left( -2 \int_{\rd}^{\re} \kappa \dr \right). \label{eq_q2} 
\end{align}
The term $q_1$ represents the coupling strength between the two g-mode cavities. Similarly, $q_2$ measures the intensity of the coupling between the outer g-mode cavity of the p-mode cavity.

One can check from Eq. \ref{eq_3cav} that the results of \cite{shibahashi79} are recovered in the cases where the coupling is negligible between the propagation cavities. For instance, if we assume that $q_1=0$, i.e. that the external g-mode cavity is uncoupled to the inner g-mode cavity, the solutions of Eq. \ref{eq_3cav} correspond to 
\begin{numcases}{}
\cot\tgun = 0 \\
\cot\tgdeux \tan\tetap = q_2. \label{eq_2cav_g2p}
\end{numcases}
The first case corresponds to the condition for pure g modes trapped in the inner cavity. The second case is identical to the condition for mixed modes given in Eq. \ref{eq_g2p}, and to the one obtained by \cite{shibahashi79} (their Eq. 31). If $q_1$ is indeed negligibly small for red giants undergoing a He subflash, then their oscillation spectrum is expected to be identical to that of regular clump giants, except for the observed period spacing ($\Delta\Pi_2$), which can be different. However, if the coupling strength $q_1$ is large enough (this is quantified in the following), the observed oscillation spectrum can become much denser because it also includes the eigenfrequencies of the inner g$_1$ cavity (modified, owing to the coupling). 

Similarly, if $q_2=0$, i.e., if no coupling exists between the internal g-mode cavities and the acoustic cavity, then Eq. \ref{eq_3cav} reduces to
\begin{numcases}{}
\tan\tetap = 0 \\
\cot\tgun \cot\tgdeux = q_1.
\end{numcases}
One thus recovers both the frequencies of pure p modes, and the frequencies of mixed modes between the g$_1$ and g$_2$ cavities. 

\subsection{Expressions of phase terms $\tgun$, $\tgdeux$, $\tetap$ \label{sect_phases}}


Eq. \ref{eq_3cav} can be used to obtain asymptotic frequencies of oscillation modes, provided the expressions of the terms $\tetap$, $\tgun$, $\tgdeux$, $q_1$, and $q_2$ are specified. 
For this purpose, we followed \cite{mosser12a}, who inserted asymptotic expressions for the frequencies of pure p and g modes inside the phase terms $\tetap$ and $\theta_{\rm g}$. We can thus approximate the phase $\tetap$ for pure $l=1$ p modes by the expression
\begin{equation}
\tetap \approx \pi \left[ \frac{\nu}{\Delta\nu} -\frac{1}{2} - \varepsilon_{\rm p} \right],
\end{equation}
where $\Delta\nu$ is the asymptotic frequency separation of consecutive p modes (the so-called ``large separation'') and $\varepsilon_{\rm p}$ is a phase offset. This ensures that when Eq. \ref{eq_ppure} is satisfied, we recover $\nu_{n,l=1} = \left(n+1/2+\varepsilon_{\rm p}\right)\Delta\nu$, which is the first-order asymptotic expression for high-order $l=1$ p modes. We note that this expression can be extended to include higher-order terms of the asymptotic development (\citealt{mosser12a}). However, in our case, there is no need to add this complexity to our development.


Similarly, the phases $\tgun$ and $\tgdeux$  can be approximated by
\begin{align}
\tgun &  \approx \pi \left( \frac{1}{\Delta\Pi_1 \nu} - \varepsilon_{\rm g_1} \right), \\
\tgdeux &  \approx \pi \left( \frac{1}{\Delta\Pi_2 \nu} - \varepsilon_{\rm g_2} \right).
\end{align}
These expressions ensure that theoretical pure g$_1$ and g$_2$ modes of degree $l=1$ are regularly spaced in period, with period spacings $\Delta\Pi_1$ and $\Delta\Pi_2$, respectively. The quantities $\varepsilon_{\rm g_1}$ and $\varepsilon_{\rm g_2}$ are phase shifts.

\subsection{Estimates of oscillation mode parameters \label{sect_mode_param}}

The pattern of the oscillation modes depends most critically on five parameters: the large separation of pure acoustic modes $\Delta\nu$, the period spacings of pure $l=1$ gravity modes in the g$_1$ and g$_2$ cavities ($\Delta\Pi_1$ and $\Delta\Pi_2$), and the coupling intensities $q_1$ and $q_2$. To estimate the mode pattern that can be expected in giants experiencing a He-core subflash, we hereafter describe the variations in these seismic parameters during the second subflash of the 1.7-$M_\odot$ \mesa\ model that was introduced in Sect. \ref{sect_1Dmodel}. The evolution is similar for other He subflashes and for other stellar masses (within the mass range where stars undergo a He flash). Values of the seismic parameters are given in Table \ref{tab_param_JWKB} for three different stages during the He subflash: at the beginning of the subflash (model 1), at an intermediate point (model 2) and at the end of the subflash (model 3). When it is relevant, we describe the way the global seismic parameters are modified when different prescriptions are used to model the boundaries of the convective regions associated to He-burning.

\subsubsection{Large separation of acoustic modes \label{sect_deltanu_JWKB}}

We obtained first estimates of the mean large separation of p modes $\Delta\nu$ for our models by using the asymptotic relation
\begin{equation}
\Delta\nu \approx \left( 2\int_0^R \frac{\hbox{d}r}{c_{\rm s}} \right)^{-1}.
\end{equation}
The large separation was found to vary between $4.5\,\mu$Hz during the subflash and $5.6\,\mu$Hz between consecutive subflashes. We also estimated the expected frequency of maximum power of the oscillations $\nu_{\rm max}$ by assuming that it scales as the acoustic cut-off frequency, i.e. that $\nu_{\rm max}\propto M R^{-2} T_{\rm eff}^{-1/2}$ (see Table \ref{tab_param_JWKB}). This assumption is supported by observations (e.g., \citealt{kjeldsen95a}, \citealt{stello08}) and has received theoretical justification (\citealt{belkacem11}). 

To first-order the asymptotic frequencies of radial modes can be written as $\nu_{n,l=0} \approx (n+\varepsilon_{\rm p})\Delta\nu$. We refined our estimates of the asymptotic large separation of p modes $\Delta\nu$ by fitting this expression to the frequencies of the radial modes of each model, which were numerically computed for each model using the oscillation code \adipls. The fits were performed over the interval $\nu_{\rm max}\pm3\Delta\nu$, using our preliminary estimates of $\Delta\nu$ and $\nu_{\rm max}$. This interval corresponds to the frequency range over which modes are detected in core-He-burning giants (\citealt{deheuvels15}). For red giants going through He subflashes, the frequency range of detected modes should be similar because the envelope properties of these stars are close to those of red clump stars. We give in Table \ref{tab_param_JWKB} the asymptotic large separations of acoustic modes that we obtained for models 1, 2, and 3 from our fits.


\subsubsection{Period spacing of gravity modes \label{sect_deltapi_JWKB}}

\begin{figure}
\begin{center}
\includegraphics[width=9cm]{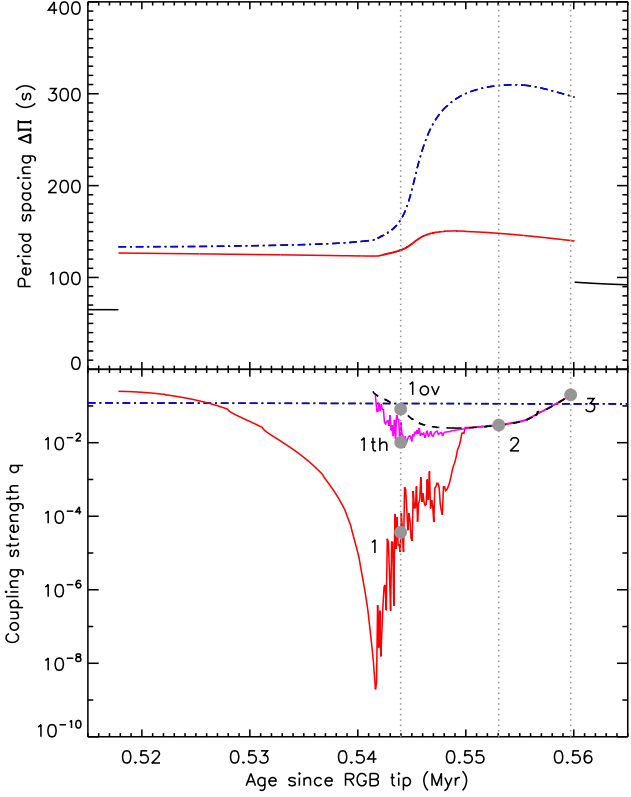}
\end{center}
\caption{Variations in global seismic parameters during the second subflash of a 1.7-$M_\odot$ model. The vertical gray dotted lines indicate the three evolutionary stages at which global seismic parameters are provided in Table \ref{tab_param_JWKB}. \textbf{Top panel:}  Asymptotic period spacing of dipolar modes in the \gun\ cavity ($\Delta\Pi_1$, red solid line) and in the \gdeux\ cavity ($\Delta\Pi_2$, blue dot-dashed line) as a function of time elapsed since the tip of the RGB. The black curve shows the period spacing of the single g-mode cavity outside of the He subflash. \textbf{Bottom panel:} Variations in the coupling strength between the \gun\ and \gdeux\ cavities ($q_1$, red solid line for Schwarzschild criterion, purple line for Ledoux criterion with thermohaline mixing, and black dashed line for Schwarzschild criterion with overshooting, see text for more details) and between the \gdeux\ and the p-mode cavities ($q_2$, blue dot-dashed line). 
\label{fig_deltapi_q}}
\end{figure}

Estimates of $\Delta\Pi_1$ and $\Delta\Pi_2$ were obtained from their asymptotic expressions
\begin{align}
\Delta\Pi_1 & \approx \pi^2\sqrt{2} \left( \int_{\ra}^{\rb} \frac{N}{r} \,\hbox{d}r \right)^{-1} \label{eq_dp1} \\
\Delta\Pi_2 & \approx \pi^2\sqrt{2} \left( \int_{\rc}^{\rd} \frac{N}{r} \,\hbox{d}r \right)^{-1} \label{eq_dp2},
\end{align}
where the turning points $\ra$, $\rb$, $\rc$, and $\rd$ were taken as those of a wave with a frequency of $\nu_{\rm max}$. The variations in the asymptotic period spacings $\Delta\Pi_1$ and $\Delta\Pi_2$ during a  He subflash are shown in the top panel of Fig. \ref{fig_deltapi_q}. When using the bare Schwarzschild criterion, the splitting of the g-mode cavity occurs earlier (about 0.518 Myr after the tip of the RGB, compared to about 0.541 for other prescriptions). This preliminary phase corresponds to the development of an evanescent zone ($N^2<0$) induced by the negative $\mu$-gradient, but stable according to the Schwarzschild criterion. This situation is known to be unstable and thermohaline mixing, if efficient enough, is expected to smooth out the $\mu$-gradient. When it is included following the prescription of \cite{brown13}, this first phase of the subflash is suppressed (see Fig. \ref{fig_deltapi_q}). Values of $\Delta\Pi_1$ and $\Delta\Pi_2$ for models 1, 2, and 3 (whose ages are indicated by vertical dotted lines in Fig. \ref{fig_deltapi_q}) are given in Table \ref{tab_param_JWKB}. They are nearly insensitive to the choice of the criterion for the onset of convection.

As mentioned in Sect. \ref{sect_1Dmodel}, most of the \gdeux-dominated modes should be detectable, so that the period spacing $\Delta\Pi_2$ should be measurable, provided the oscillation spectrum of the star can be deciphered. If thermohaline mixing is efficient, $\Delta\Pi_2$ lies between 200 and 300 s during most of the subflash, and is thus comparable to the period spacing of regular clump stars. For inefficient mixing, $\Delta\Pi_2$ is significantly smaller than the period spacings of clump stars (as shown by \citealt{vrard16}, very few red clump giants have measured period spacings below 200 s) during the first half of the subflash.

The period spacing of the \gun\ cavity is systematically smaller than $\Delta\Pi_2$ but it is of the same order of magnitude. This already suggests that the oscillation spectra generated by both g-mode cavities taken separately should have a similar density. 


\subsubsection{Coupling strength \label{sect_coupling}}

\begin{figure}
\begin{center}
\includegraphics[width=9cm]{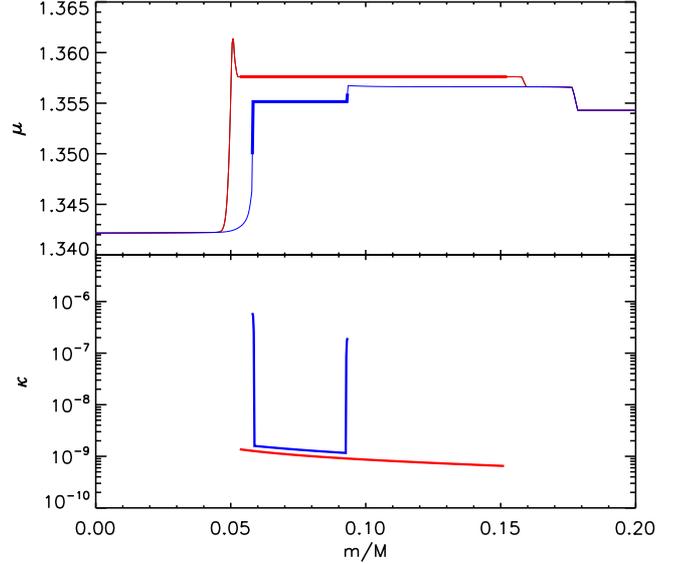}
\end{center}
\caption{Profiles of the mean molecular weight $\mu$ (\textbf{top}) and the function $\kappa$ (\textbf{bottom}) in the core of Model 1 (blue curves) and Model 2 (red curves). In the top panel, thick lines indicate the evanescent zone separating the \gun- and \gdeux-cavities.
\label{fig_evol_mu}}
\end{figure}

One critical point was to estimate the coupling strength between the cavities. 
The expressions of the coupling strength given by Eq. \ref{eq_q1} and \ref{eq_q2} have been shown to be valid only in the case of a weak coupling, i.e., when the evanescent zone that separates the cavities is thick enough (\citealt{takata16}). For thin evanescent zones, these expressions tend to underestimate the coupling strength (\citealt{takata16})
Calculating coupling strength for He-flashing giants in the general case would require to adapt the formalism of \cite{takata16} to the special case of three mode cavities, which represents considerable work. Here, we chose to use Eq. \ref{eq_q1} and \ref{eq_q2} to obtain first estimates of the coupling strength between the cavities, which provide insights about the main contributors to the coupling intensities. 

We plugged the expression of $\kappa$ (Eq. \ref{eq_kr}) into Eq. \ref{eq_q1} and \ref{eq_q2} to estimate $q_1$ and $q_2$, whose variations during the subflash are shown in Fig. \ref{fig_deltapi_q}. For the coupling intensity between the outer g-mode cavity and the p-mode cavity, we found values of $q_2$ varying between 0.11 and 0.15, with minimal values reached during the subflashes and maximal values between subflashes. 
The coupling strength $q_1$ between the two g-mode cavities depends on the modeling of the boundaries of the convective region in which He is burnt. 
\begin{itemize}
\item When using the bare Schwarzschild criterion, a thin evanescent zone develops at the beginning of the subflash, owing to the negative $\nabla\mu$, as mentioned in Sect. \ref{sect_deltapi_JWKB}. As this zone extends, the coupling $q_1$ decreases by several orders of magnitude. The existence of this preliminary phase is uncertain since it is prone to thermohaline mixing, which could smooth out $\mu$-gradients. About 0.542 Myr after the RGB tip, a part of the evanescent zone becomes convectively unstable according to the Schwarzschild criterion and mixing occurs in this region. The coupling strength $q_1$ becomes dominated by the contribution of the narrow regions adjacent to the convective shell in which sharp negative $\mu$-gradients develop. This is clearly seen in the profile of $\kappa$ for Model 1 shown in Fig. \ref{fig_evol_mu}. Since the turning points $\rb$ and $\rc$, which delimit the evanescent zone, coincide with the regions of sharp $\mu$-gradients, small variations in $\rb$ and $\rc$ can induce large changes in the value of $q_1$. This explains at least partly the spiky features of the curve showing the variations in $q_1$ during the first half of the subflash (Fig. \ref{fig_deltapi_q}). Small numerical errors in the computation of the \vaisala\ frequency within the \mesa\ code could also contribute to this spiky behavior. The $\mu$-gradients are progressively smoothed out at both edges, causing $q_1$ to increase, as can be seen in Fig. \ref{fig_deltapi_q}. Indeed, the carbon abundance in the convective shell increases, which reduces the jump in the mean molecular weight at the outer edge of the shell. Also, the layers below the convective shell are progressively heated and they start burning helium into carbon, which increases their mean molecular weight (see top panel of Fig. \ref{fig_evol_mu}). About 0.549 Myr after the RGB tip, the $\mu$-gradients beyond both edges of the convective shell become positive. The regions of varying chemical composition outside the convective shell no longer contribute to the coupling intensity $q_1$ (see Fig. \ref{fig_evol_mu}), which slowly increases as the shell shrinks toward the end of the subflash.
\item Using the Ledoux criterion with thermohaline mixing following the prescription of \cite{brown13}, the beginning of the subflash is delayed, as explained above. When convection is triggered (around 0.542 Myr after the RGB tip), the regions with negative $\mu$-gradients on both sides of the convective shell are part of the evanescent zone. However, thermohaline mixing smoothes out these gradients and $q_1$ is much higher than with the Schwarzschild criterion. When the $\mu$-gradients become positive outside the convective shell, $q_1$ no longer depends on them and becomes indistinguishable from the bare-Schwarzschild case.
\item When a mild amount of overshooting is added, the regions of sharp $\mu$-gradients are pushed away to the boundaries of the overshooting layers. In these regions, the temperature gradient is equal to the radiative temperature gradient, which is significantly below the adiabatic temperature gradient. As a result, $N^2>0$ even when $\nabla_\mu$ is negative. The coupling strength $q_1$ thus remains independent of the $\mu$-gradients outside the convective region throughout the subflash and depends only on the thickness of the shell.
\end{itemize} 

The largest differences between the values of $q_1$ obtained with the different prescriptions occur when convection starts in the He-burning shell (model 1). In the following sections, we refer to the model computed with the Schwarzschild criterion as `model 1', the model computed with the Ledoux criterion and thermohaline mixing as `model 1th', and the model computed with overshooting as `model 1ov'. For models 2 and 3, $q_1$ is independent of the adopted criterion.

 Our asymptotic calculations predict that the coupling strength $q_1$ should be above $10^{-2}$ at least during the largest part of the subflash, and potentially throughout the subflash if mixing is included (thermohaline mixing or overshooting). However, these conclusions should be taken with care. Indeed, at the beginning and in the end of the subflash, the evanescent region is very thin and the formalism of \cite{takata16} should be used. Also, around the edges of the evanescent zone, the mean molecular weight varies over length scales that can be smaller than the wavelengths of the modes, which violates the main assumption behind the JWKB approximation. The intensity of the coupling $q_1$ between the two g-mode cavities thus remains uncertain. To address the question of the mode visibilities in Sect. \ref{sect_height}, we thus considered three scenarios: a weak coupling ($q_1 = 10^{-3}$), an intermediate coupling ($q_1 = 10^{-2}$) and a strong coupling ($q_1 = 10^{-1}$). The asymptotic values of $q_1$ are compared to the results of full numerical calculations of oscillation frequencies in Sect. \ref{sect_numerical}.

\begin{table}
  \begin{center}
  \caption{Seismic parameters of the reference model (1.7-$M_\odot$ model during the second He-core subflash) calculated using their asymptotic expressions (see text). \label{tab_param_JWKB}}
\begin{tabular}{c c c c c c}
\hline \hline
\T \B Model & 1 & 1th & 1ov & 2 & 3 \\
\hline
\T $\nu_{\rm max}$ ($\mu$Hz) & 36.8 & 36.6 & 36.6 & 36.2 & 35.8  \\
$\Delta\nu$ ($\mu$Hz) & 4.7 & 4.5 & 4.6 & 4.5 & 4.5 \\
$\Delta\Pi_1$ (s)  & 130.6 & 129.3 & 129.8 & 148.8 & 143.6 \\
$\Delta\Pi_2$  (s) & 165.0 & 164.8 & 164.0 & 307.0 & 299.9  \\
$q_1$ & $2\times10^{-6}$ & 0.01 & 0.10 & 0.03 & 0.10 \\
\B $q_2$ & 0.12 & 0.12 & 0.12 & 0.11 & 0.11 \\
 \hline 
\end{tabular}
\end{center}
\end{table}

\subsection{Synthetic spectrum \label{sect_freq_JWKB}}


\begin{figure}
\begin{center}
\includegraphics[width=9cm]{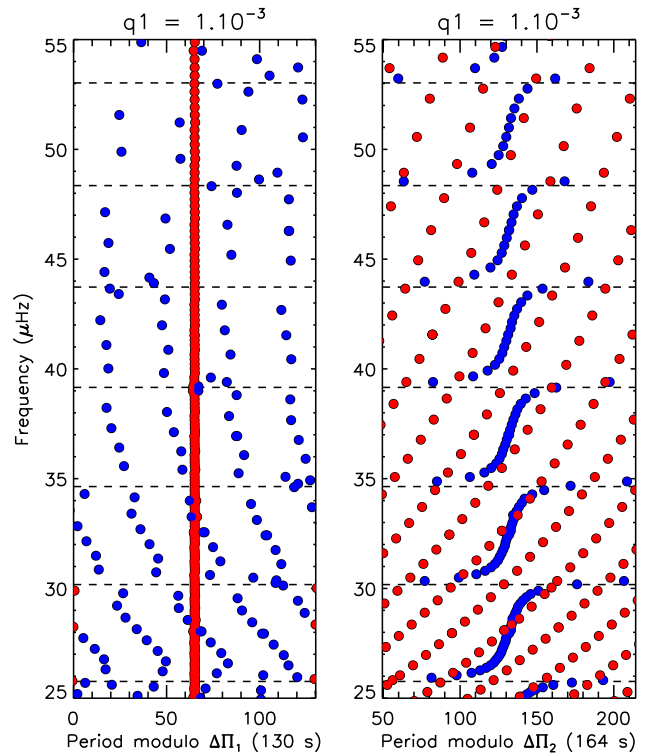}
\end{center}
\caption{Period \'echelle diagrams of asymptotic oscillation modes obtained by solving Eq. \ref{eq_3cav} using the parameters given in Table \ref{tab_param_JWKB} and considering a weak coupling between the g-mode cavities ($q_1=10^{-3}$). These diagrams were obtained by folding the mode periods using either $\Delta\Pi_1$ (left) or $\Delta\Pi_2$ (right). The horizontal dashed lines indicate the location of theoretical pure $l=1$ p modes. Modes trapped predominantly in the g$_1$ cavity ($I_{\rm g_1}/I > 0.5$) are filled in red, and modes trapped mainly in the g$_2$ cavity ($I_{\rm g_2}/I > 0.5$) are filled in blue (mode trapping is estimated based on ratios of inertia, see Sect. \ref{sect_inertia}). Other modes are left blank. \label{fig_echelle_JWKB_period}}
\end{figure}

\begin{figure}
\begin{center}
\includegraphics[width=9cm]{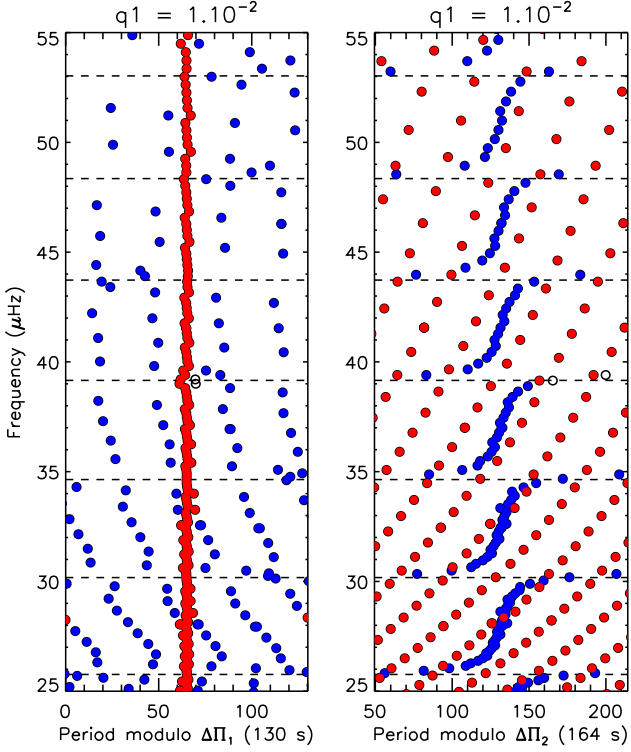}
\end{center}
\caption{Same as Fig. \ref{fig_echelle_JWKB_period} for an intermediate coupling between the g-mode cavities ($q_1=10^{-2}$). 
\label{fig_echelle_JWKB_period2}}
\end{figure}

\begin{figure}
\begin{center}
\includegraphics[width=9cm]{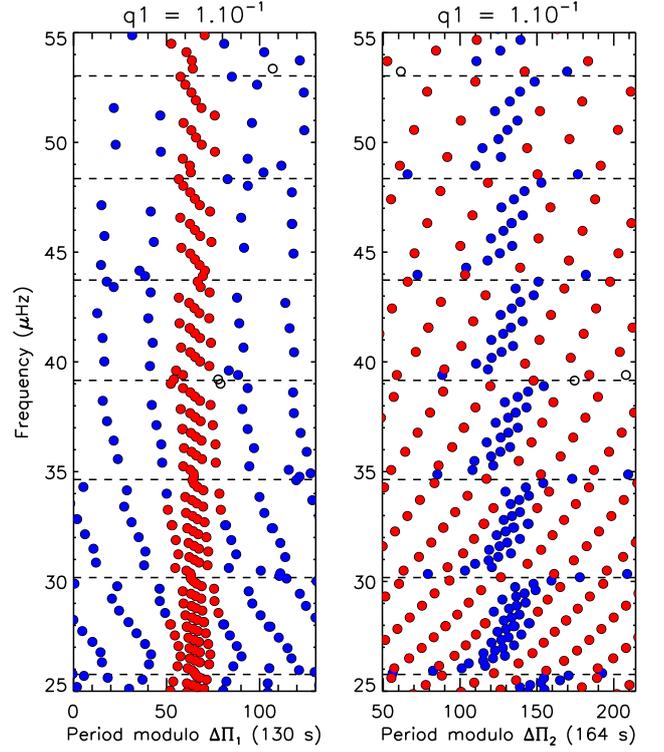}
\end{center}
\caption{Same as Fig. \ref{fig_echelle_JWKB_period} for a strong coupling between the g-mode cavities ($q_1=10^{-1}$). 
\label{fig_echelle_JWKB_period3}}
\end{figure}

We then had all the ingredients to build asymptotic oscillation spectra for our reference models. For this purpose, we solved Eq. \ref{eq_3cav} using a Newton-Raphson algorithm for frequencies in the frequency range $\nu_{\rm max}\pm3\Delta\nu$, which corresponds to the expected frequency of maximum power of the oscillations. The oscillation spectra were calculated following this procedure for the three different stages of the He subflash that are represented by vertical dotted lines in Fig. \ref{fig_deltapi_q}. As already hinted above, we found that the coupling strength $q_1$ between the g-mode cavities plays a key role in shaping the oscillation spectra of models in the He subflash. The remaining seismic global parameters do not qualitatively alter the pattern of the oscillation spectra. Consequently, we found it more instructive to show in this section the results corresponding to the seismic parameters of a single model (we chose those of model 1) and to study the influence of varying $q_1$ on this particular model.


As expected, the obtained oscillation spectrum is quite dense. It should be comprised essentially of modes trapped in the g$_1$ or g$_2$ cavities, with a few p-dominated modes. As a first step to separate these modes, we used the obtained mode periods to build period \'echelle diagrams\footnote{Period \'echelle diagrams are built by cutting the spectrum of mode periods in chunks of size $\Delta\Pi$ and piling them up. This has the consequence of regrouping modes that are regularly spaced in period by $\Delta\Pi$ as vertical ridges in the period \'echelle diagram.} folded using either $\Delta\Pi_1$ (Fig. \ref{fig_echelle_JWKB_period} to \ref{fig_echelle_JWKB_period3}, left panels) or $\Delta\Pi_2$ (Fig. \ref{fig_echelle_JWKB_period} to \ref{fig_echelle_JWKB_period3}, right panels). These \'echelle diagrams correspond to three different scenarios concerning the coupling strength between the g-mode cavities: a weak ($q_1=10^{-3}$, Fig. \ref{fig_echelle_JWKB_period}), an intermediate ($q_1=10^{-2}$, Fig. \ref{fig_echelle_JWKB_period2}), and a strong coupling ($q_1=10^{-1}$, Fig. \ref{fig_echelle_JWKB_period3}). Decreasing the coupling intensity $q_1$ below $10^{-3}$ does not significantly modify the oscillation spectrum.

The \'echelle diagrams folded with $\Delta\Pi_1$ show a clear straight vertical ridge, which corresponds to modes trapped predominantly in the inner g$_1$ cavity. The departures from verticality in the case of a strong coupling $q_1$ are caused by avoided crossings with g$_2$ modes. The remaining modes are mainly trapped either in the g$_2$ cavity or in the acoustic-mode cavity. These modes are clearly identified in the \'echelle diagrams folded with $\Delta\Pi_2$ (right panels of Fig. \ref{fig_echelle_JWKB_period} to \ref{fig_echelle_JWKB_period3}) because they regroup along S-shaped ridges, which are the result of avoided crossings between modes trapped in the g$_2$ cavity and acoustic modes that are regularly spaced in frequency (horizontal dashed lines in Fig. \ref{fig_echelle_JWKB_period} to \ref{fig_echelle_JWKB_period3}). These features are reminiscent of the period \'echelle diagrams that are obtained for regular red giants (see e.g. \citealt{mosser12a}). The widening of the S-shaped ridges that is observed in the strong coupling case ($q_1=0.1$) is caused by the avoided crossings between modes trapped in the g$_1$-cavity and modes trapped in the \gdeux-cavity.


\subsection{Mode trapping \label{sect_inertia}}

To go further, one needs to know where the different modes are trapped inside the star. For this purpose, it is necessary to estimate the mode inertias $I$ (which are directly related to the kinetic energy of the modes) inside each of the three cavities. 
We thus denote as $I_{\rm g_1}$, $I_{\rm g_2}$, and $I_{\rm p}$ the inertias in each cavity. \cite{goupil13}  proposed an approximate expression for the mode inertias based on JWKB analysis in the case of two cavities. We extended their work to the case of three cavities using the asymptotic development presented in Appendix \ref{app_JWKB}. The details of the calculation are provided in Appendix \ref{app_ratios}. We were able to estimate the fraction of the total energy of the mode that is trapped in each of the three cavities, which is given by $I_{\rm g_1}/I$, $I_{\rm g_2}/I$, and $I_{\rm p}/I$. The expressions for these quantities are given by Eq. \ref{eq_IpvsI} to \ref{eq_Ig2vsI} (or alternatively by Eq. \ref{eq_IpvsI_alt} to \ref{eq_Ig2vsI_alt} when the former equations are singular, see Appendix \ref{app_ratios} for more details). 


We calculated these quantities for each of the asymptotic oscillation modes of our reference model. To illustrate the results, we highlighted the modes for which at least 50\% of the mode kinetic energy is trapped within the g$_1$ cavity ($I_{\rm g_1}/I>0.5$) as filled red circles in Fig. \ref{fig_echelle_JWKB_period} to \ref{fig_echelle_JWKB_period3}. They indeed correspond to the modes that form the vertical ridges in the period \'echelle diagrams folded with $\Delta\Pi_1$. Similarly, the modes that are trapped mainly in the g$_2$ cavity ($I_{\rm g_2}/I>0.5$) are shown as filled blue circles in Fig. \ref{fig_echelle_JWKB_period} to \ref{fig_echelle_JWKB_period3}. They correspond to the S-shaped ridges in the period \'echelle diagrams folded with $\Delta\Pi_2$.

\begin{figure}
\begin{center}
\includegraphics[width=9cm]{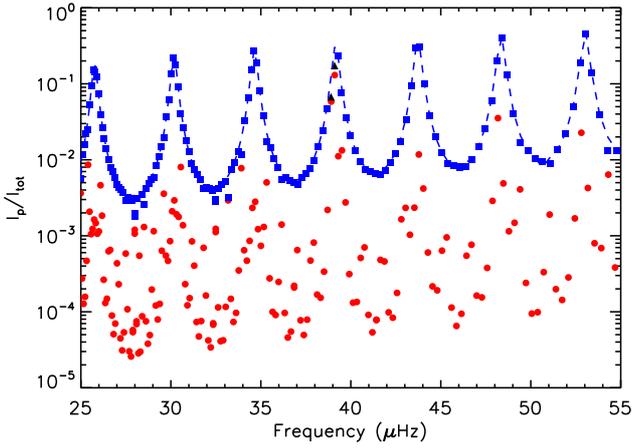}
\end{center}
\caption{Fractional contribution of the acoustic cavity to the mode inertia (and thus to the mode kinetic energy) using the global seismic parameters of model 1 and assuming an intermediate coupling between the two g-mode cavities ($q_1=10^{-2}$). Modes that have more than 50\% of the energy trapped in the  g$_1$ (resp. g$_2$) cavity are shown as filled red circles (resp. blue squares). Other modes are shown as filled black triangles. The dashed blue curve indicates the ratios of inertia $I_{\rm p}/I_{\rm tot}$ that would be obtained for mixed modes trapped only in the g$_2$ and p mode cavities. 
\label{fig_inertia_JWKB}}
\end{figure}

The ratios of inertia $I_{\rm p}/I$ correspond to the fractional contribution of the acoustic cavity to the kinetic energy of the modes. They are often used as a proxy for mode visibilities. Indeed, for modes with large inertias, this quantity approximately corresponds to the ratio between the height of the considered $l=1$ mode in the power spectrum and the height of a pure p mode, provided the effects of radiative damping are neglected (e.g., \citealt{grosjean14}). This statement is justified in Sect. \ref{sect_height}, where the effects of radiative damping are also addressed. Fig. \ref{fig_inertia_JWKB} shows the ratios of inertia $I_{\rm p}/I$ for the modes computed using Eq. \ref{eq_IpvsI} in the case of an intermediate coupling strength between the two g-mode cavities ($q_1 = 10^{-2}$). For comparison, we overplotted the ratios of inertia $I_{\rm p}/I$ that would be obtained if the modes were trapped only in the g$_2$- and p-mode cavities (blue dashed curve). The regularly-spaced maxima of this curve correspond to the location of theoretical pure p modes. The shape of this curve has been extensively studied (e.g. \citealt{mosser12a}, \citealt{goupil13}, \citealt{mosser15}). The modes that lie along this blue dashed curve correspond to modes that are g$_2$-p dominated, while those that lie outside of it have a non-negligible part of their energy trapped within the innermost g$_1$ cavity. For most of the latter modes, the acoustic cavity contributes very little to the mode energy, which makes them unlikely to be observed. However, Fig. \ref{fig_inertia_JWKB} shows the existence of modes that are trapped mainly in the g$_1$ cavity ($I_{\rm g_1}/I>0.5$) and yet have non-negligible contribution of the acoustic cavity to the mode energy. The number of such modes naturally increases with the intensity of the coupling $q_1$ between the two g-mode cavities. These modes, which are located in the neighborhood of theoretical pure $l=1$ p modes have better chances of being detected. To estimate this quantitatively, we evaluated their damping rate in Sect. \ref{sect_height}.

\begin{figure}
\begin{center}
\includegraphics[width=9cm]{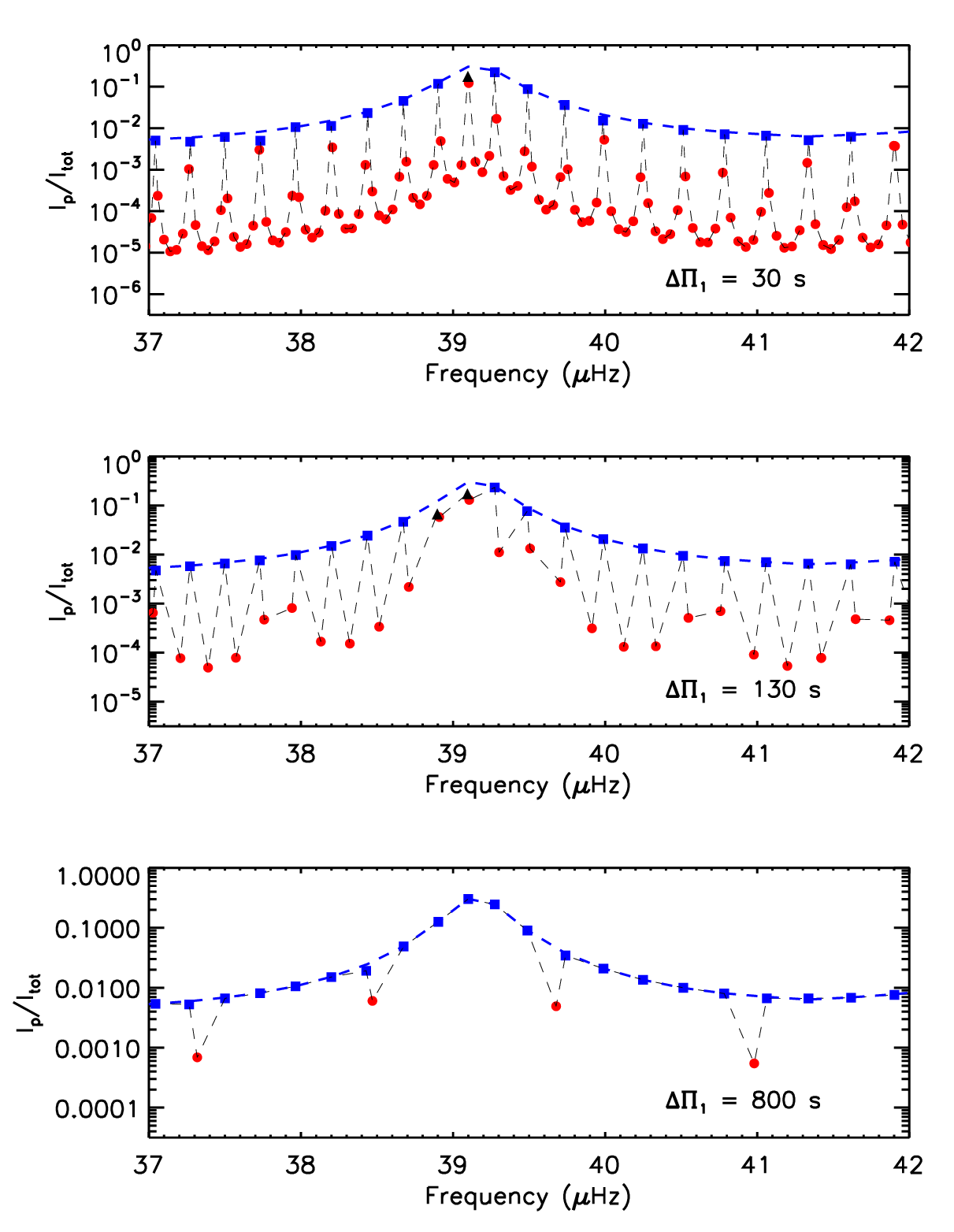}
\end{center}
\caption{Same as Fig. \ref{fig_inertia_JWKB} but assuming values of the asymptotic period spacing $\Delta\Pi_1$ ranging from 30 to 800 s. To guide the eye, the ratio $I_{\rm p}/I$ for consecutive modes have been linked by the black dashed curve.
\label{fig_inertia_alt}}
\end{figure}

We note that the variations in the ratios of inertia obtained for g$_1$-dominated modes (filled red circles in Fig. \ref{fig_inertia_JWKB}) seem quite erratic. The reason for this is the similarity of the period spacings in the two g-mode cavities. This causes the oscillation spectra of both g-mode cavities to have roughly the same density. As a consequence, consecutive modes are trapped alternately in the g$_1$ and g$_2$ cavities and the curve representing the ratios of inertia has a sawtooth behavior. To make this point clearer, we also calculated mode frequencies and ratios of inertia by artificially changing the period spacing in the g$_1$ cavity to either a much lower value than the period spacing in the g$_1$ cavity (we took $\Delta\Pi_1=30$ s) or a much larger value ($\Delta\Pi_1=800$ s). The obtained ratios of inertia $I_{\rm p}/I$ are shown in Fig. \ref{fig_inertia_alt}. In the former case, the oscillation spectrum of the g$_1$ cavity is much denser than that of the g$_2$ cavity. As a result, between each pair of g$_2$-dominated mode, $I_{\rm p}/I$ shows a dip, reminiscent of what is obtained in the case of the trapping between an acoustic cavity and a single g-mode cavity. In the latter case ($\Delta\Pi_1=800$ s), the spectrum of the g$_1$ cavity is much sparser. $I_{\rm p}/I$ follows closely the curve corresponding to the case where modes are trapped only in the g$_2$- and p-mode cavities (blue dashed curve) with sharp dips corresponding to modes trapped in the g$_1$ cavity.


\subsection{Mode heights \label{sect_height}}

One critical point in our study is to predict which modes are likely to be observed among the rich oscillation spectrum obtained from JWKB analysis. The relevant quantity to study this important question is the expected height of the modes in the power spectrum. This quantity depends on the damping rates of the modes and on the duration of the observations. The damping rate $\eta$ of a mode is given by the expression
\begin{equation}
\eta = -\frac{\int \hbox{d}W}{2\omega^2 I}
\label{eq_eta}
\end{equation}
where $\int \hbox{d}W$ is the work performed by the gas during one oscillation cycle, $\omega$ is the mode angular frequency, and $I=\int_0^M |\vect{\xi}|^2\,\hbox{d}m$ is the mode inertia. As proposed by \cite{grosjean14}, the work integral can be separated into two contributions: the radiative damping in regions corresponding to both g-mode cavities ($W_{\rm g_1}$ and $W_{\rm g_2}$, respectively) and the outer non-adiabatic part of the convective envelope ($W_{\rm e}$), so that
\begin{equation}
\int \hbox{d}W = W_{\rm g_1} + W_{\rm g_2} + W_{\rm e}
\end{equation}
We examine both contributions in the following.

\subsubsection{Surface damping rates \label{sect_etap}}

For low-degree modes, the displacement is nearly radial in the outer convective region, where the modes are excited. As a consequence, we can assume that the work $W_{\rm e}$ is identical for $l=1$ mixed modes and for radial modes. For radial modes, radiative damping in the interior can be safely neglected, and the damping rate is thus given by:
\begin{equation}
\eta_{\rm p} = \frac{-W_{\rm e}}{2\omega^2I_{\rm p}}  \label{eq_we} 
\end{equation}
The damping rates of radial modes due to non-adiabatic effects can be estimated observationally using the linewidths of these modes (the mode linewidth $\Gamma$ is related to the damping rate through the relation $\eta=\pi\Gamma$). The radial modes of red clump stars have a typical width of the order of 0.1 to 0.2 $\mu$Hz (\citealt{deheuvels15}, \citealt{mosser17b}). This provides a damping rate of about $\eta_{\rm p}\approx 5\times 10^{-7}$ s$^{-1}$. This value is in broad agreement with the scaling law $\eta\propto T_{\rm eff}^{10.8} g^{-0.3}$ prescribed by \cite{belkacem12} on the basis of numerical simulations that include time-dependent convection. We therefore adopted the quoted value of $\eta_{\rm p}$ as an estimate of the damping rate undergone by the modes in the outer convective region. Eq. \ref{eq_we} was then used to estimate the work $W_{\rm e}$ induced by non-adiabatic effects, which we assumed to hold also for $l=1$ modes, as mentioned above.

\subsubsection{Radiative damping}

In the g-mode cavities, the high density induces a very high \vaisala\ frequency, which can produce radiative damping. It is important to quantify this effect in our case, because if the modes that are trapped significantly in the innermost g$_1$ cavity are too severely damped, they might have very low visibilities, leaving us little chance to detect them. Simple expressions for radiative damping have been obtained in the asymptotic limit (\citealt{dziembowski77}, \citealt{godart09}), which we have used here. In our case, we must separate the contributions from both g-mode cavities. In the innermost g$_1$ cavity, the radial displacement can be expressed as
\begin{equation}
\frac{\xi_r}{r} \sim \frac{a}{\sqrt{\pi}} \frac{\left[l(l+1)\right]^{1/4}}{\sqrt{\rho\omega r^5 N}} \cos\left( \int_{\ra}^r k_r \dr - \frac{\pi}{4} \right)
\end{equation}
where we have used Eq. \ref{eq_vg1}. Following the development of \cite{godart09}, we found that the contribution of the g$_1$ cavity to the work integral is given by
\begin{equation}
-W_{\rm g_1} \simeq \frac{a^2}{\pi} \mathcal{J}(\ra,\rb) \label{eq_wg1}
\end{equation}
where
\begin{equation}
\mathcal{J}(\ra,\rb) \equiv \frac{\left[l(l+1)\right]^{3/2}}{2\omega^3} \int_{\ra}^{\rb} \frac{\nabla_{\rm ad}-\nabla}{\nabla}\frac{\nabla_{\rm ad}NgL}{p r^5}\,\hbox{d}r \\
\end{equation}
where $\nabla_{\rm ad}$ and $\nabla$ are the adiabatic and real temperature gradients, $g$ is the local gravity, $L$ the local luminosity, and $p$ is the pressure. Similarly, 
using the asymptotic expression of $\xi_r$ in the g$_2$ cavity (see Appendix \ref{app_JWKB}), we can obtain the following expression for the contribution of the \gdeux\ cavity to the radiative damping:
\begin{equation}
-W_{\rm g_2} \simeq \frac{c'^2+d'^2}{\pi} \mathcal{J}(\rc,\rd) \label{eq_wg2} \\
\end{equation}
The integrals $\mathcal{J}(\ra,\rb)$ and $\mathcal{J}(\rc,\rd)$ were calculated using our reference model.

\subsubsection{Total damping rates}

By plugging Eq. \ref{eq_we}, \ref{eq_wg1}, and \ref{eq_wg2} into Eq. \ref{eq_eta}, we obtain the total damping rate of g modes $\eta_{\rm mix}$, which can be expressed as follows
\begin{align}
\eta_{\rm mix} & = - \frac{W_{\rm g_1} + W_{\rm g_2} + W_{\rm e}}{2\omega^2 I} \\
& =  \frac{a^2 \mathcal{J}(\ra,\rb) + (c'^2+d'^2) \mathcal{J}(\rc,\rd)}{2\pi\omega^2I} + \frac{I_{\rm p}}{I}\eta_{\rm p} \\
 & = \frac{1}{4\pi} \left[ \frac{I_{\rm g_1}}{I} \frac{\mathcal{J}(\ra,\rb)}{\theta_{\rm g_1}} + \frac{I_{\rm g_2}}{I} \frac{\mathcal{J}(\rc,\rd)}{\theta_{\rm g_2}} \right] + \frac{I_{\rm p}}{I} \eta_{\rm p}
 \label{eq_etag}
\end{align}
where Eq. \ref{eq_Ig2} and \ref{eq_Ig1} have been used. The ratios of inertia can be estimated using Eq. \ref{eq_IpvsI},  \ref{eq_Ig1vsI}, and \ref{eq_Ig2vsI} and the phases $\theta_{\rm g_i}$ are approximated by $\pi/(\nu\Delta\Pi_i)$, as explained in Appendix \ref{app_ratios}.
Fig. \ref{fig_etag} shows the damping rates obtained for the modes of a model undergoing a He-subflash. 

\begin{figure}
\begin{center}
\includegraphics[width=9cm]{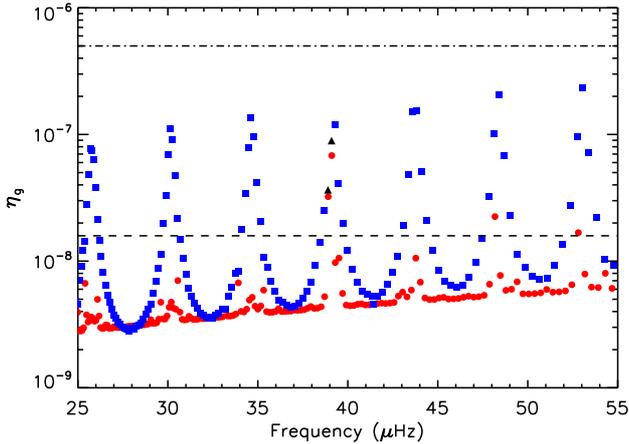}
\end{center}
\caption{Damping rates $\eta_{\rm mix}$ of mixed modes obtained from a JWKB analysis using the global seismic parameters of model 1 and assuming an intermediate coupling between the two g-mode cavities ($q_1=10^{-2}$). The symbols are the same as in Fig. \ref{fig_inertia_JWKB}. The horizontal dash-dotted line indicates the damping rates of p modes inferred from seismic observations of typical clump stars (see text). The threshold $\eta_{\rm lim}=2/T_{\rm obs}$ below which modes are unresolved is represented by the horizontal dashed line (we assumed a duration of $T_{\rm obs}=4$ yr, corresponding to the longest datasets of \kepler). 
\label{fig_etag}}
\end{figure}

\subsubsection{Height ratios}

The height of a mixed mode in the power spectrum depends on whether or not the mode is resolved (e.g., \citealt{dupret09}). A mode is resolved if its lifetime $\tau$ is shorter than $T_{\rm obs}/2$, where $T_{\rm obs}$ is the duration of observations, which is approximately 4 years for the longest datasets available with \kepler\ data. Since the mode lifetime corresponds to the inverse of the damping rate, we can deduce a limit damping rate $\eta_{\rm lim}=2/T_{\rm obs}$, below which the modes are unresolved. This limit has been overplotted in Fig. \ref{fig_etag} in the case of the full \kepler\ datasets. We can see that a large fraction of the modes that are mainly trapped in the g$_2$ and p-cavities are expected to be resolved when using \kepler\ data. On the contrary, most of the modes trapped in the g$_1$ cavity are expected to be unresolved, with the exception of a few modes in the vicinity of theoretical pure $l=1$ p modes.

In the case of resolved mixed modes, \cite{grosjean14} have shown that the ratio between their height and that of a p-type mode is given by
\begin{equation}
\frac{H_{\rm mix}^{\rm res}}{H_{\rm p}} = \left(1+\frac{W_{\rm g_1}+W_{\rm g_2}}{W_{\rm e}}\right)^{-2}
\label{eq_res}
\end{equation}
This expression involves the ratio between the radiative damping in the g-mode cavities and the damping due to non-adiabatic effects in the outer convective envelope, which can be deduced from Eq. \ref{eq_we}, \ref{eq_wg1}, and \ref{eq_wg2}:
\begin{equation}
\frac{W_{\rm g_1}+W_{\rm g_2}}{W_{\rm e}} = \frac{1}{4\pi\eta_{\rm p}\theta_{\rm p}} \left[ \frac{a^2}{c''^2}\mathcal{J}(\ra,\rb) + \frac{c'^2+d'^2}{c''^2}\mathcal{J}(\rc,\rd) \right]
\label{eq_work_ratio_JWKB}
\end{equation}

For unresolved modes, the height ratio is given by
\begin{equation}
\frac{H_{\rm mix}^{\rm unres}}{H_{\rm p}} = \left(1+\frac{W_{\rm g_1}+W_{\rm g_2}}{W_{\rm e}}\right)^{-1} \frac{I_{\rm p}}{I}\eta_{\rm p}\frac{T_{\rm obs}}{2}
\label{eq_unres}
\end{equation}
as shown by \cite{grosjean14}. We note that if the effects of radiative damping are neglected, the height ratio between an $l=1$ mixed mode and a pure p mode is inversely proportional to the ratio of inertia of the two modes, as was stated in Sect. \ref{sect_inertia}.

\begin{figure}
\begin{center}
\includegraphics[width=9cm]{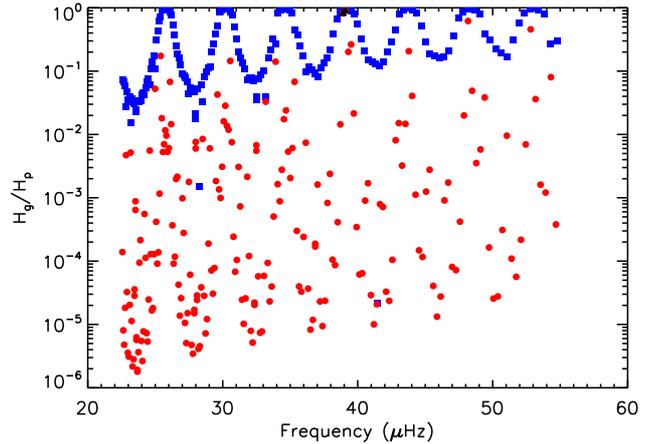}
\end{center}
\caption{Height ratios between $l=1$ mixed modes and theoretical $l=1$ pure p mode obtained from a JWKB analysis using the global seismic parameters of model 1 and assuming an intermediate coupling between the two g-mode cavities ($q_1=10^{-2}$). These ratios were calculated using either Eq. \ref{eq_res} (if the mode is resolved) or Eq. \ref{eq_unres} (otherwise).
\label{fig_height}}
\end{figure}

For each mode, we used Eq. \ref{eq_etag} to determine whether the mode is resolved or not. We then estimated the height ratio using either Eq. \ref{eq_res} (if the mode is resolved) or Eq. \ref{eq_unres} (otherwise). The results are plotted in Fig. \ref{fig_height}. This figure shows that for intermediate coupling intensities between the two g-mode cavities ($q_1=10^{-2}$ in Fig. \ref{fig_height}), several modes that are mainly trapped in the g$_1$ cavity are expected to have heights comparable to those of radial modes in the power spectrum, which means that they should be detectable.

\subsection{What would the spectrum of a giant in a He-core subflash look like? \label{sect_stretch_JWKB}}

\begin{figure*}
\begin{center}
\includegraphics[width=5cm]{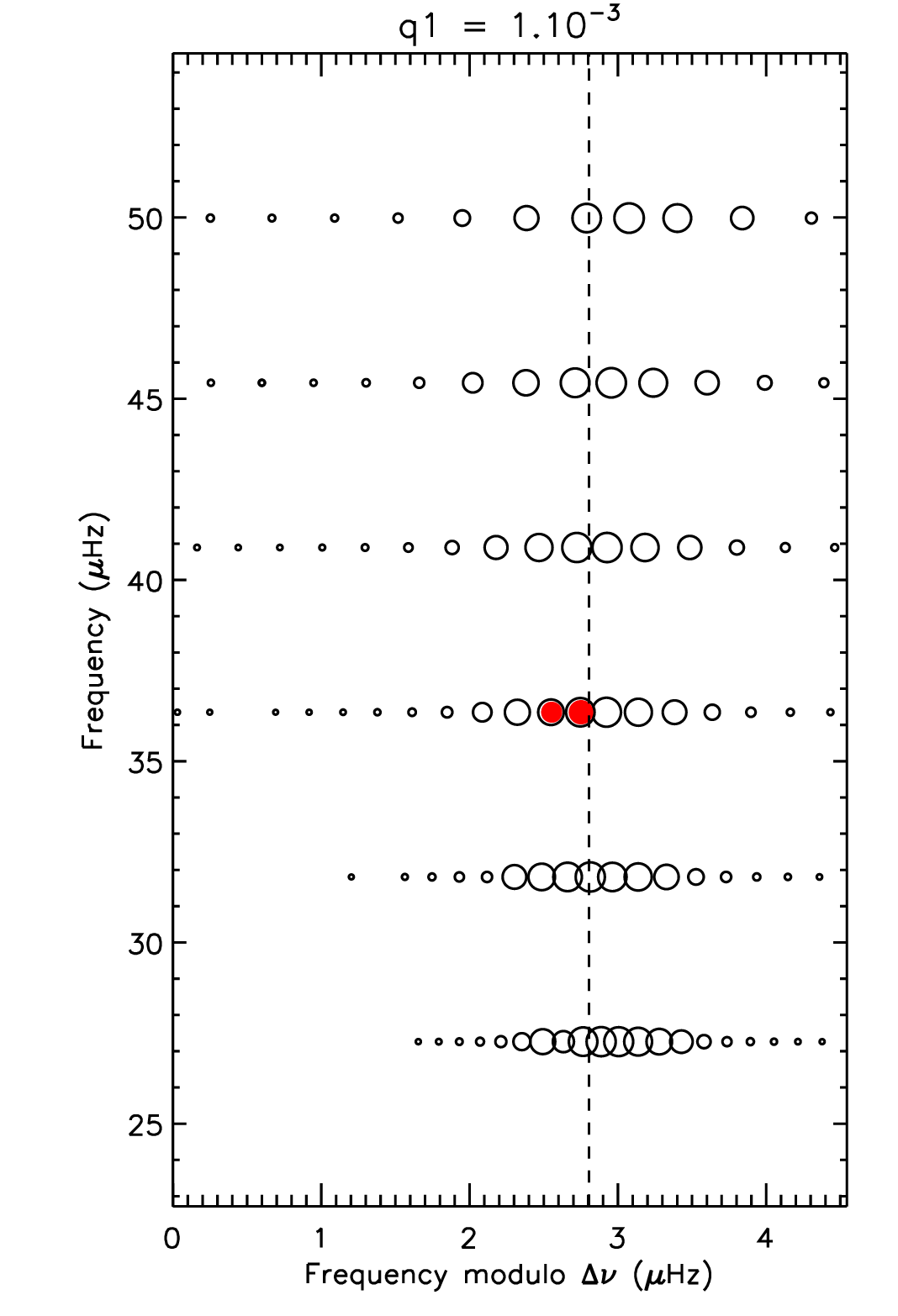}
\includegraphics[width=5cm]{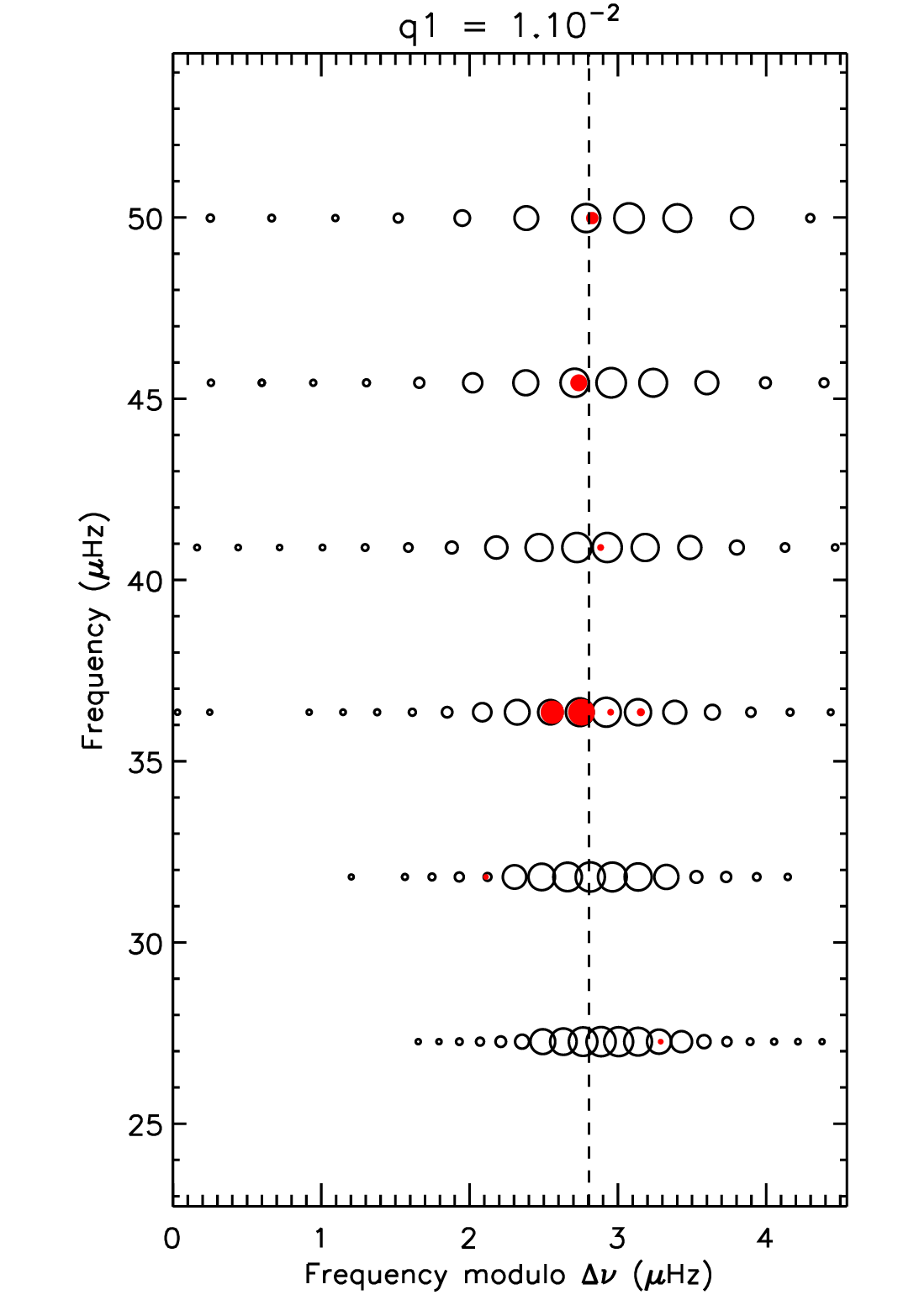}
\includegraphics[width=5cm]{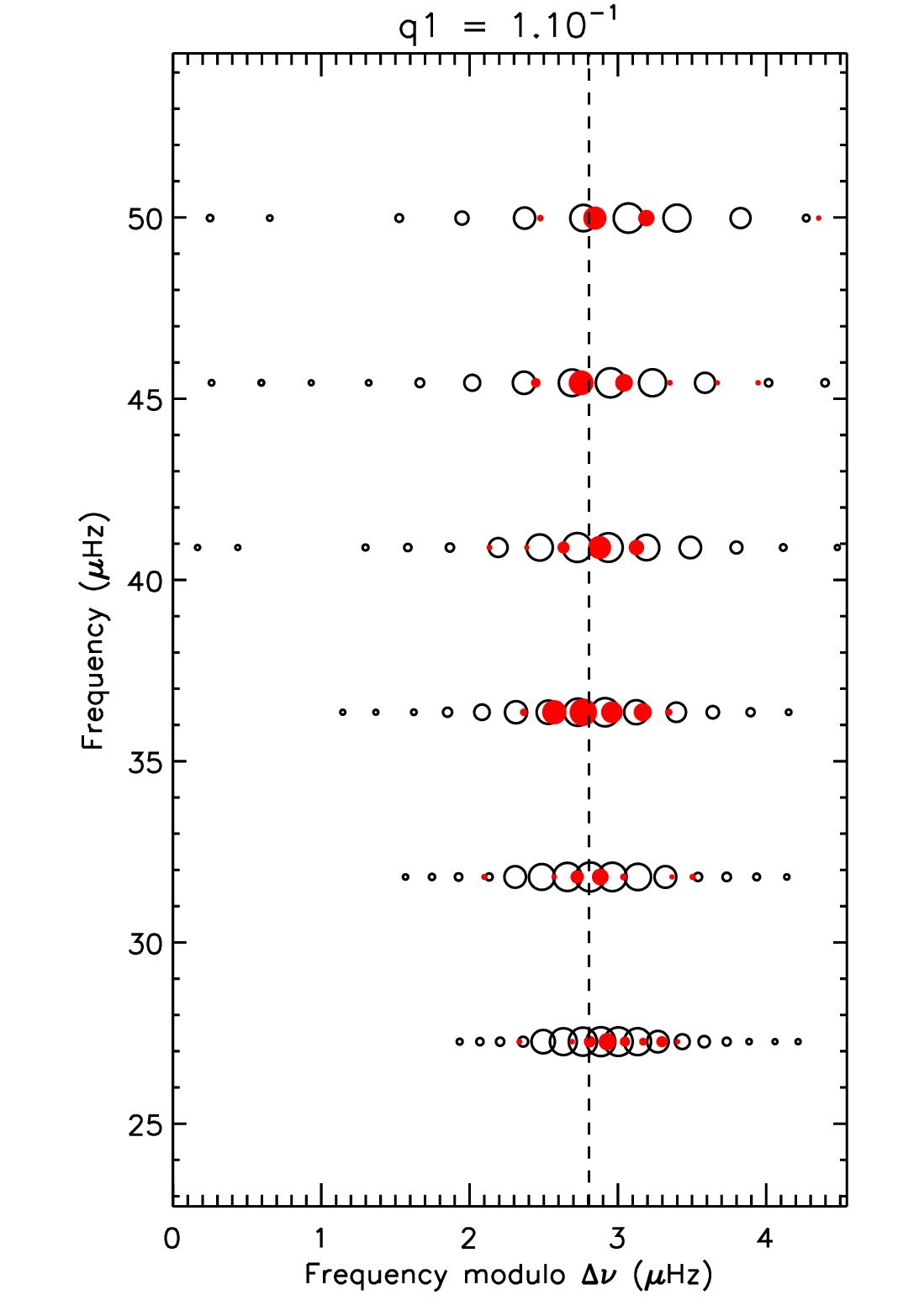}
\includegraphics[width=5cm]{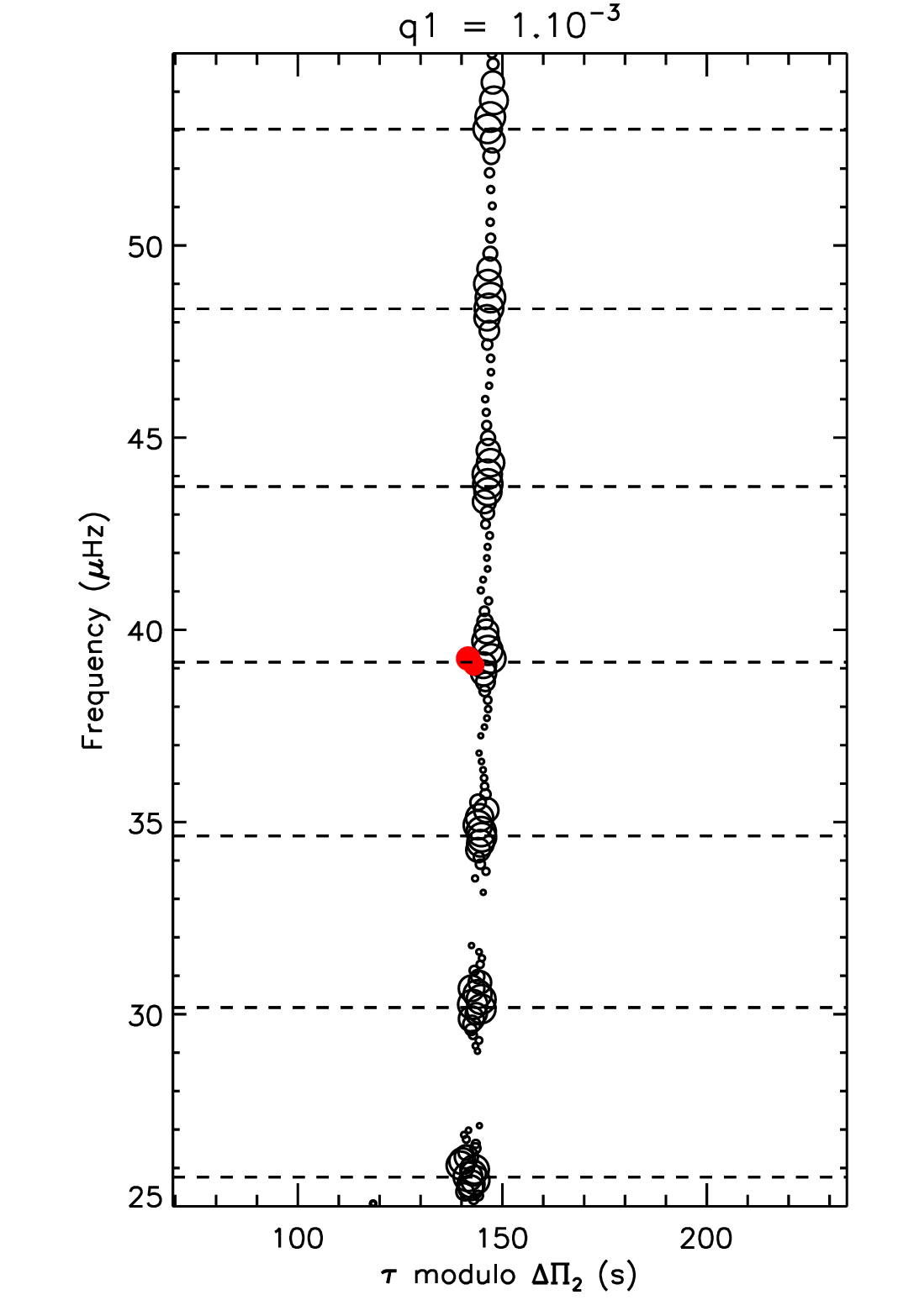}
\includegraphics[width=5cm]{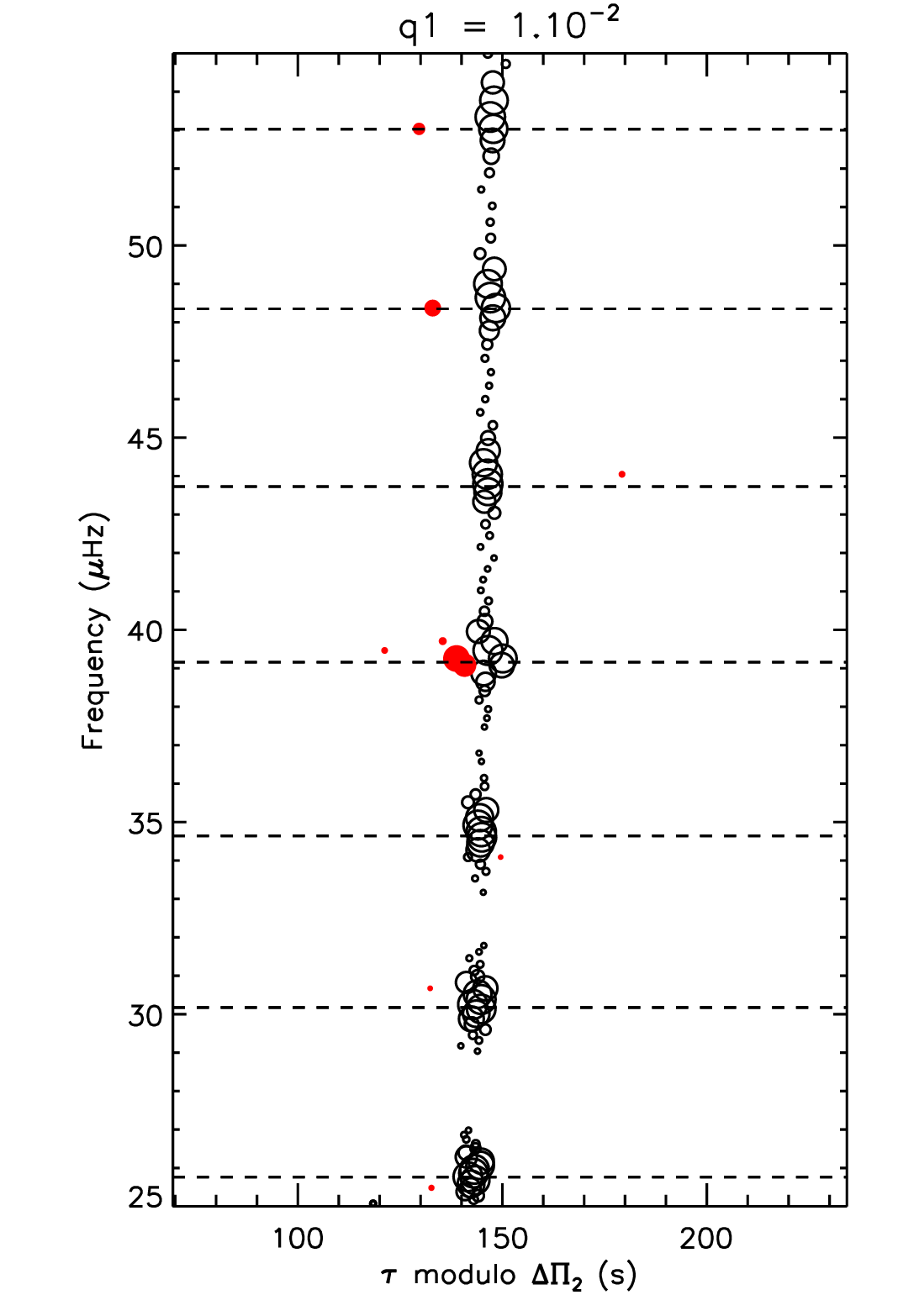}
\includegraphics[width=5cm]{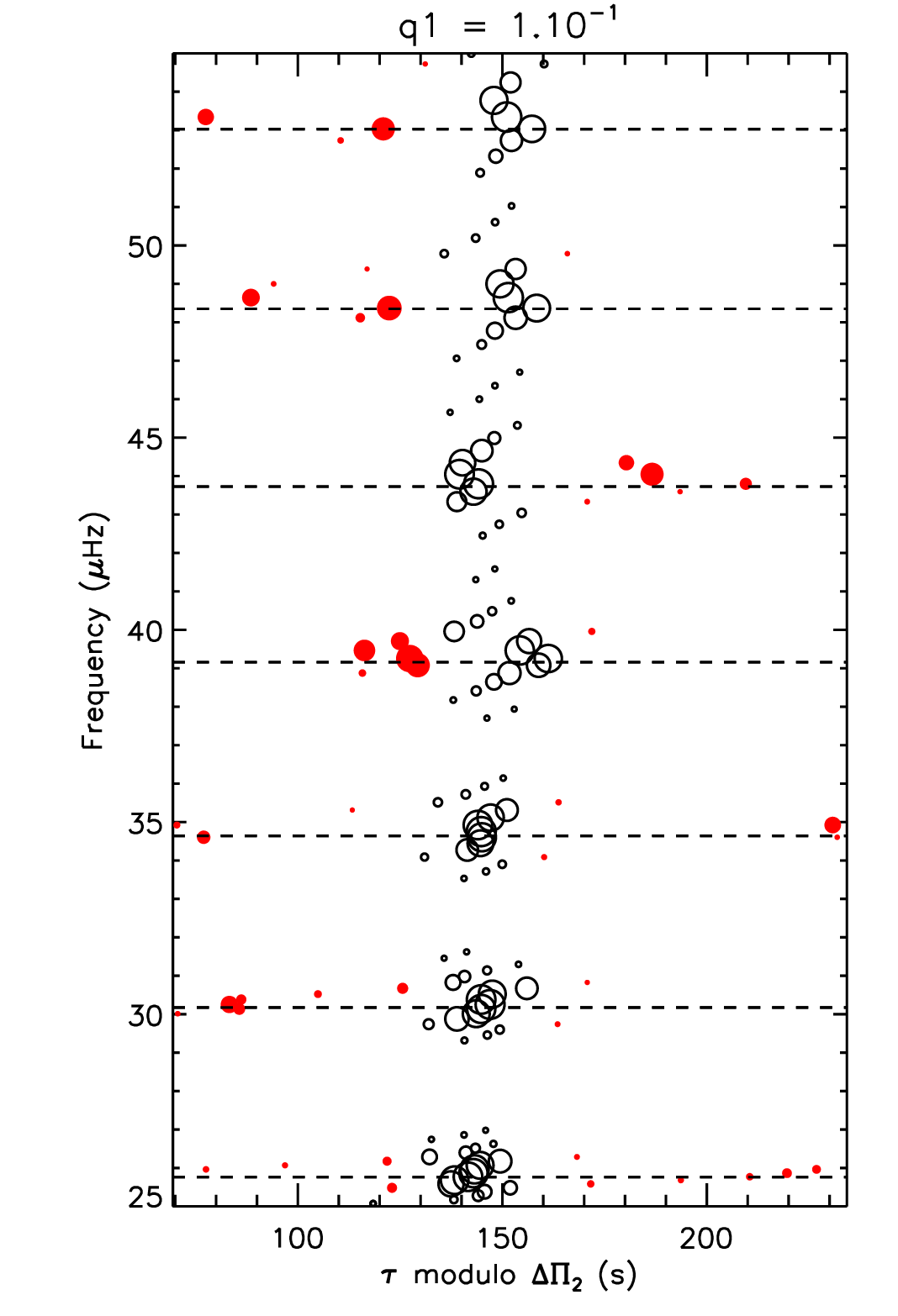}
\end{center}
\caption{\'Echelle diagrams of the oscillation spectrum of model 1 obtained using the JWKB approximation, and assuming different coupling intensities between the g-mode cavities ($q_1=10^{-3}$, $10^{-2}$, $10^{-1}$ from left to right). The size of the symbols indicates the expected height of each mode. For clarity, only the modes with a height corresponding to at least 10\% of the height of a pure p mode are shown. Modes colored in red have at least 50\% of their total energy trapped in the innermost g$_1$ cavity (i.e. $I_{\rm g_1}/I > 0.5$). \textbf{Top panels:} Frequency \'echelle diagrams folded with the the large separation of acoustic modes $\Delta\nu$. \textbf{Bottom panels:} Stretched period \'echelle diagrams built following \cite{mosser15} (see text). 
\label{fig_echelle_ratio}}
\end{figure*}

We can now address the main question of interest, i.e., what would the oscillation spectrum of a red giant undergoing a He subflash look like? To answer this question, we represented the oscillation spectrum obtained from our JWKB analysis by weighting the modes according to their expected heights, based on the results of Sect. \ref{sect_height}. 

We used two types of representation. First, we showed the oscillation spectrum by building a classical \'echelle diagram folded using the large separation of acoustic modes, $\Delta\nu=4.7\,\mu$Hz in our case (left panels of Fig. \ref{fig_echelle_ratio}). For clarity, we plotted only the modes whose expected heights correspond to at least 10\% of the height of a pure p mode. For a weak coupling between the two g-mode cavities ($q_1<10^{-2}$), only a few modes that are trapped in the g$_1$ cavity have significant predicted heights and the spectrum is indistinguishable from the one that we would obtain if we ignored the g$_1$ cavity altogether. For stronger coupling intensities ($q_1\gtrsim10^{-2}$), the oscillation spectrum corresponds to the spectrum of \gdeux- and p-dominated modes, complemented with additional modes that are trapped mainly in the g$_1$ cavity and have expected heights comparable to those of pure p modes. Quite expectedly, these additional modes are located in the neighborhood of theoretical pure acoustic $l=1$ modes (see Fig. \ref{fig_echelle_ratio}).

In order to identify these additional modes, the so-called ``stretched'' period \'echelle diagram that was proposed by \cite{mosser15} is a particularly helpful representation. We here briefly describe the construction of these diagrams, referring the reader to their paper for more details. In the usual case of two cavities, the period spacings of $l=1$ mixed modes vary, which results in  S-shaped ridges in the period \'echelle diagrams (see for instance blue circles in Fig. \ref{fig_echelle_JWKB_period} to \ref{fig_echelle_JWKB_period3}). \cite{mosser15} showed that the period spacings between $l=1$ mixed modes can be related to the $\zeta$ function, which is defined as the ratio of the kinetic energy of modes in the g-mode cavity versus their total kinetic energy. Conveniently, theoretical expressions for the $\zeta$ function have been obtained from JWKB analysis (\citealt{goupil13}, \citealt{deheuvels15}). \cite{mosser15} proposed to express the oscillation spectrum as a function of a modified (``stretched'') period $\tau$ instead of the regular mode period $P$, where $\tau$ is defined by the differential equation $\hbox{d}\tau = {\hbox{d}P}/{\zeta}$. This has the consequence of forcing the oscillation modes to be regularly spaced when using the modified period $\tau$. In other words, this approach straightens the ridge of $l=1$ modes in the period \'echelle diagram. In our case, we expect the oscillation spectrum to be comprised essentially of the modes that are trapped mainly in the g$_2$ and p cavities, along with additional modes trapped in the \gun\ cavity, whose number depends on the value of $q_1$. We can thus build stretched \'echelle diagrams taking into account only the g$_2$ and p cavities. The modes that are trapped in these cavities should then regroup along a single vertical ridge. Additional modes, which are trapped mainly in the g$_1$ cavity, are not expected to lie on this ridge and they can thus be easily spotted in this type of diagram. Fig. \ref{fig_echelle_ratio} (right column) shows the results obtained for the three coupling intensities that were tested. For $q_1\gtrsim10^{-2}$, additional modes are clearly identified outside of the vertical ridge. We also note that as $q_1$ increases, the vertical ridge widens. This is caused by the avoided crossings between g$_1$ and g$_2$ modes which significantly modify the frequencies of modes trapped in the g$_2$ and p cavities.

To summarize, we found that if the coupling between the two g-mode cavities is large enough ($q_1\gtrsim10^{-2}$) during He subflashes, the observed oscillation spectrum should be comprised mostly of modes trapped in the \gdeux\ and p cavities, but it should also include additional modes with detectable heights, which are trapped mainly in the \gun\ cavity. In this case, these additional modes could efficiently be identified by plotting a stretched period \'echelle diagram. If the coupling is weaker ($q_1<10^{-2}$), the oscillation spectrum during the subflash is expected to be very similar to that of regular clump stars (except for the value of the period spacing, which could differ). But even in this case, some modes would be \gun-dominated and could be used to seismically identify stars undergoing a He subflash (see Sect. \ref{sect_discussion}).

\section{Numerical calculation of mode frequencies during a He subflash \label{sect_numerical}}

To complement the asymptotic analysis performed in Sect. \ref{sect_JWKB}, we numerically solved the full equations of stellar oscillations for the \mesa\ model that was introduced in Sect. \ref{sect_1Dmodel}. Since oscillation codes have never been thoroughly tested in this evolution stage, we chose to use two different codes: the Aarhus adiabatic oscillation package \adipls\ (\citealt{adipls}) and the oscillation code \gyre\ (\citealt{gyre}). Both codes have already been adapted to calculate oscillation modes for evolved red giants. We have used the option available for both oscillation codes to reset the mesh grid in order to ensure that the eigenfunctions of the modes are correctly resolved. This requires a fine meshing in the core of the star. A comparison between the results of the two codes is presented in Appendix \ref{app_adipls_gyre}. Despite mode-to-mode differences that will need to be looked into in future works, the agreement between \adipls\ and \gyre\ is sufficient to validate the numerical calculation of mode frequencies and mode trapping during He subflashes, to the level of precision required in this study. We thus present only the results obtained using \adipls\ in this Section.

\subsection{Oscillation spectrum during a He subflash \label{sect_freq_model}}


\begin{figure*}
\begin{center}
\includegraphics[width=\textwidth]{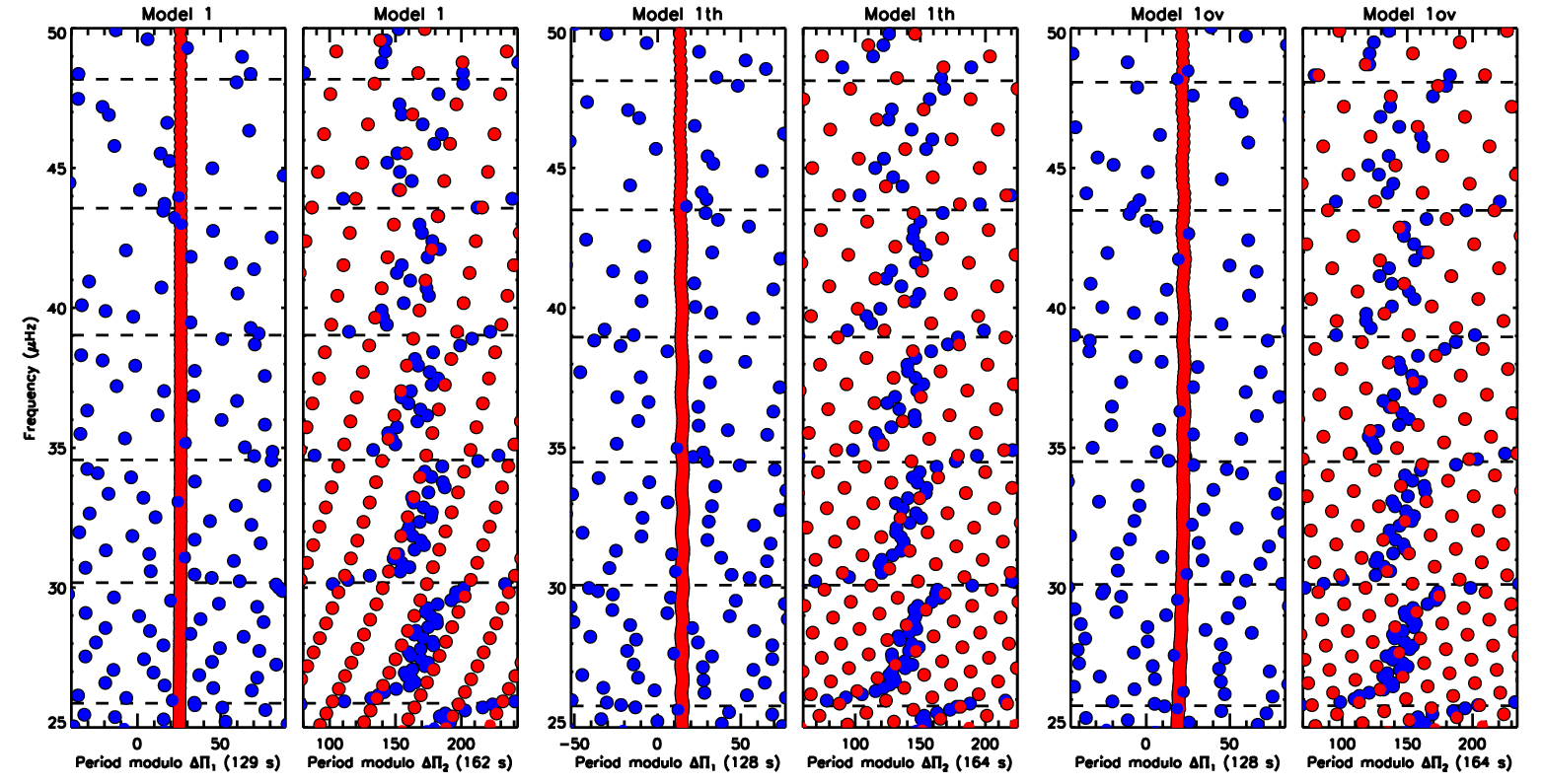}
\includegraphics[width=0.68\textwidth]{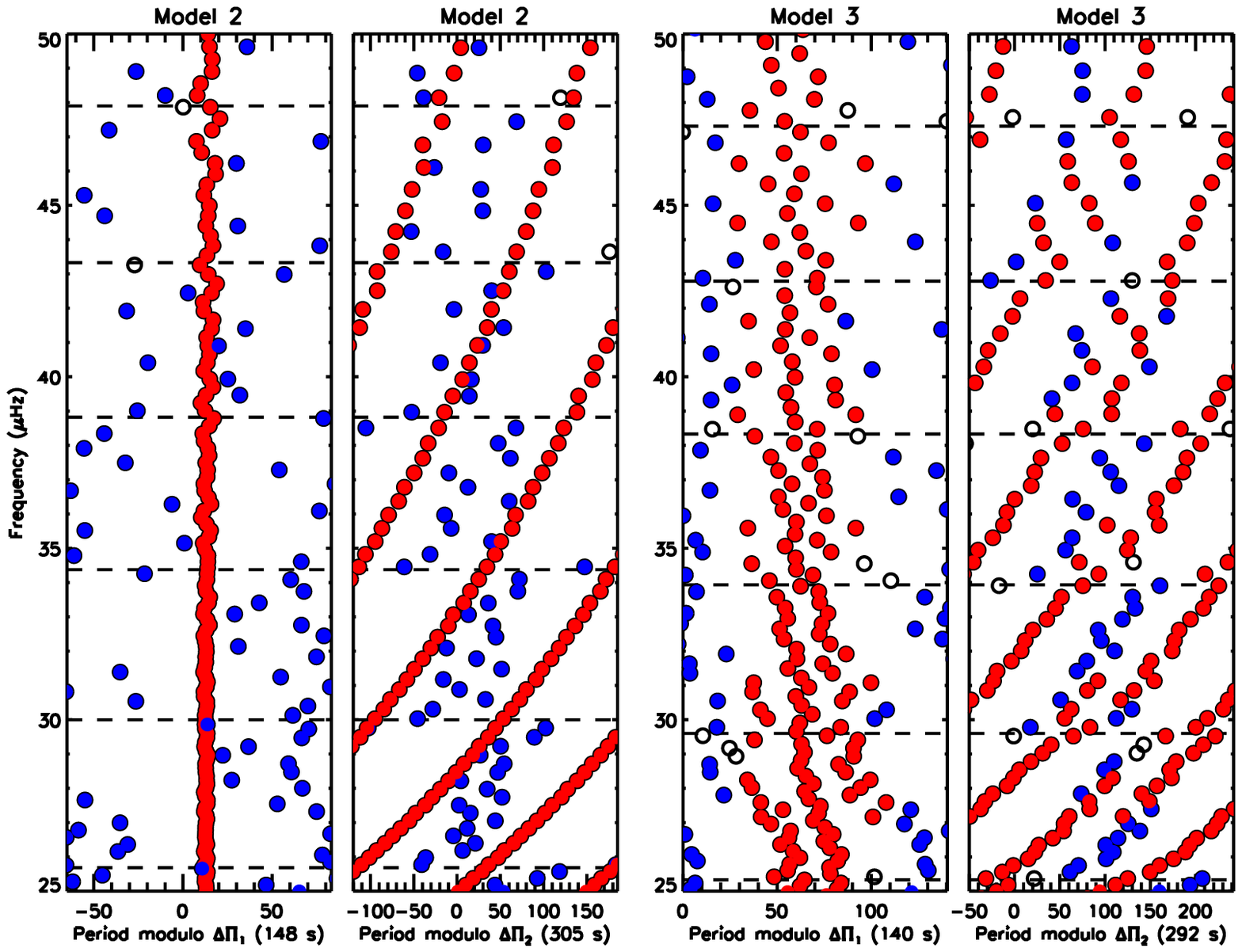}
\end{center}
\caption{Same as Fig. \ref{fig_echelle_JWKB_period} but using the oscillation modes that were computed with \adipls\ for a 1.7-$M_\odot$ \mesa\ model at several stages during the second He subflash (top left panels: model 1, top right panels: model 1ov, bottom left panels: model 2, bottom right panels: model 3, see text for description). For each model, period \'echelle diagrams were constructed using the period spacings of the g$_1$ cavity (left plots) and the g$_2$ cavity (right plots). \label{fig_echelle_adipls}}
\end{figure*}


We calculated the oscillation mode frequencies for models 1, 1th, 1ov, 2, and 3, which were presented in Sect. \ref{sect_1Dmodel}. For each model, we built period \'echelle diagrams that were folded using alternately the period spacing of the g$_1$ cavity and that of the g$_2$ cavity (Fig. \ref{fig_echelle_adipls}). The values of $\Delta\Pi_1$ and $\Delta\Pi_2$ were slightly adjusted compared to the asymptotic values obtained from Eq. \ref{eq_dp1} and \ref{eq_dp2} in order to obtain vertical ridges in the period \'echelle diagrams. For models 1, 1th, 1ov, and 2, this adjustment was of the order of 1~s or below. For model 3, which corresponds to the very end of the He subflash, the adjustment reached 8~s for $\Delta\Pi_2$. 

For full numerical solutions of the oscillation equations, estimating the trapping of the modes is straightforward because the energy of the modes in each of the three cavities can be estimated using directly the mode eigenfunctions. 
We could thus directly determine in which region each mode is predominantly trapped. Modes for which the g$_1$ (resp. \gdeux) cavity was found to contribute to more than half of the mode kinetic energy were highlighted as red (resp. blue) filled circles in Fig. \ref{fig_echelle_adipls}. 

\subsection{Comparison with calculations using the JWKB approximation}

The period \'echelle diagrams of models 1, 1th, and 1ov can be directly compared to those of Fig. \ref{fig_echelle_JWKB_period}, \ref{fig_echelle_JWKB_period2}, and \ref{fig_echelle_JWKB_period3}, which were obtained using the JWKB approximation and assuming period spacings corresponding to those of models 1 and 1ov. The first observation is that there seems to be a fairly good qualitative agreement between the predictions obtained with the JWKB approximation and the numerical results. Similarly to the results of Sect. \ref{sect_JWKB}, the \'echelle diagrams folded with $\Delta\Pi_1$ exhibit a clear vertical line and those folded with $\Delta\Pi_2$ show the typical S-shaped ridges, whose branches correspond to avoided crossings with acoustic modes. There are however, differences between the numerically computed mode frequencies and those predicted using the JWKB approximation, which are described below.

\subsubsection{Coupling between the g-mode cavities}

For model 1ov, the large value of $q_1$ (0.10, see Table \ref{tab_param_JWKB}) led us to expect a period \'echelle diagram similar to that of Fig. \ref{fig_echelle_JWKB_period3}. Instead, the oscillation spectrum of model 1ov is very similar to that of model 1, for which $q_1\sim 10^{-3}$. In particular, for model 1ov, the ridge formed by the \gun-dominated modes in the \'echelle diagram folded using $\Delta\Pi_1$ is remarkably straight. This suggests a much weaker coupling than predicted by the JWKB approximation. This discrepancy is further studied when we address the question of mode trapping in Sect. \ref{sect_trapping_adipls}. 

By contrast, for model 3, the ridge of \gun-dominated modes in the period \'echelle diagram appears significantly scattered, which is consistent with the large coupling strength that is predicted by the JWKB approximation ($q_1=0.10$, see Table \ref{tab_param_JWKB}).

\subsubsection{Effects of buoyancy glitches \label{sect_glitch}}

We also observe that the S-shaped ridges formed by \gdeux-dominated modes in the \'echelle diagrams folded using $\Delta\Pi_2$ exhibit oscillations that were not present in the \'echelle diagrams built using the JWKB approximation. These oscillations are most easily identified in the \'echelle diagrams of models 1 and 1ov for frequencies above $\sim$ 40 $\mu$Hz. For these two models, each period of the oscillation contains approximately six \gdeux-dominated modes. 

These oscillations are caused by glitches in the \vaisala\ frequency (commonly referred to as a \textit{buoyancy glitch}), which are known to have such an effect on the period spacings (see e.g. \citealt{miglio08}, \citealt{cunha15}). Fig. \ref{fig_diag_prop} shows that the \vaisala\ frequency indeed features a sharp peak in the \gdeux-cavity, which is produced by the H-burning shell. The apparent increase of the amplitude of these oscillations with mode frequency (see Fig. \ref{fig_echelle_adipls}) supports the hypothesis that the observed oscillation is the result of a buoyancy glitch. This behavior is expected for glitch-related oscillations, which arise when the equilibrium quantities vary on a length-scale comparable to or shorter than the mode wavelength. For gravity modes, the wavelength increases with increasing frequency. Thus, if we consider a sharp variation in the \vaisala\ frequency over a length-scale $\ell$, high-frequency g modes with a wavelength larger than $\ell$ feel this perturbation as a glitch, contrary to low-frequency modes with a wavelength shorter than $\ell$. This explains the increasing amplitude of the observed oscillations.

We also calculated the oscillation period that the H-burning shell is expected to produce. Its expression in terms of the radial order $n$ is given by
\begin{equation}
\Delta n =  \frac{1}{x_{\rm Hshell}},
\label{eq_period_osc}
\end{equation}
where $x_{\rm Hshell}$ is the normalized buoyancy radius H-burning shell in the \gdeux-cavity, defined as
\begin{equation}
x_{\rm Hshell} = \frac{ \displaystyle \int_{\rc}^{r_{\rm Hshell}} \frac{N}{r} \,\hbox{d}r }{ \displaystyle \int_{\rc}^{\rd} \frac{N}{r} \,\hbox{d}r  }.
\label{eq_pishell}
\end{equation}
For models 1 and 1ov, we found a normalized buoyancy radius of $x_{\rm Hshell}\approx0.82$ for the H-burning shell, which should give $\Delta n \approx 1.22$ according to Eq. \ref{eq_period_osc}. In practice, periodicities such that $\Delta n < 2$ cannot be directly observed, as stated by the Nyquist-Shannon sampling theorem. In this case, the signature of the glitch is aliased to a signal whose source would be located at a normalized buoyancy radius of $1-x_{\rm Hshell}$ (see \citealt{montgomery03}, \citealt{miglio08}). This gives an expected period of $\Delta n \approx 5.6$ for the H-burning-shell-glitch, which is in very good agreement with the visual estimate of six radial orders per period that was obtained from Fig. \ref{fig_echelle_adipls}. This confirms that the observed oscillations are indeed produced by the H-burning shell. We further discuss the implications of this oscillation in Sect.~\ref{sect_discussion}.

\subsection{Mode trapping \label{sect_trapping_adipls}}

\begin{figure}
\begin{center}
\includegraphics[width=8.5cm]{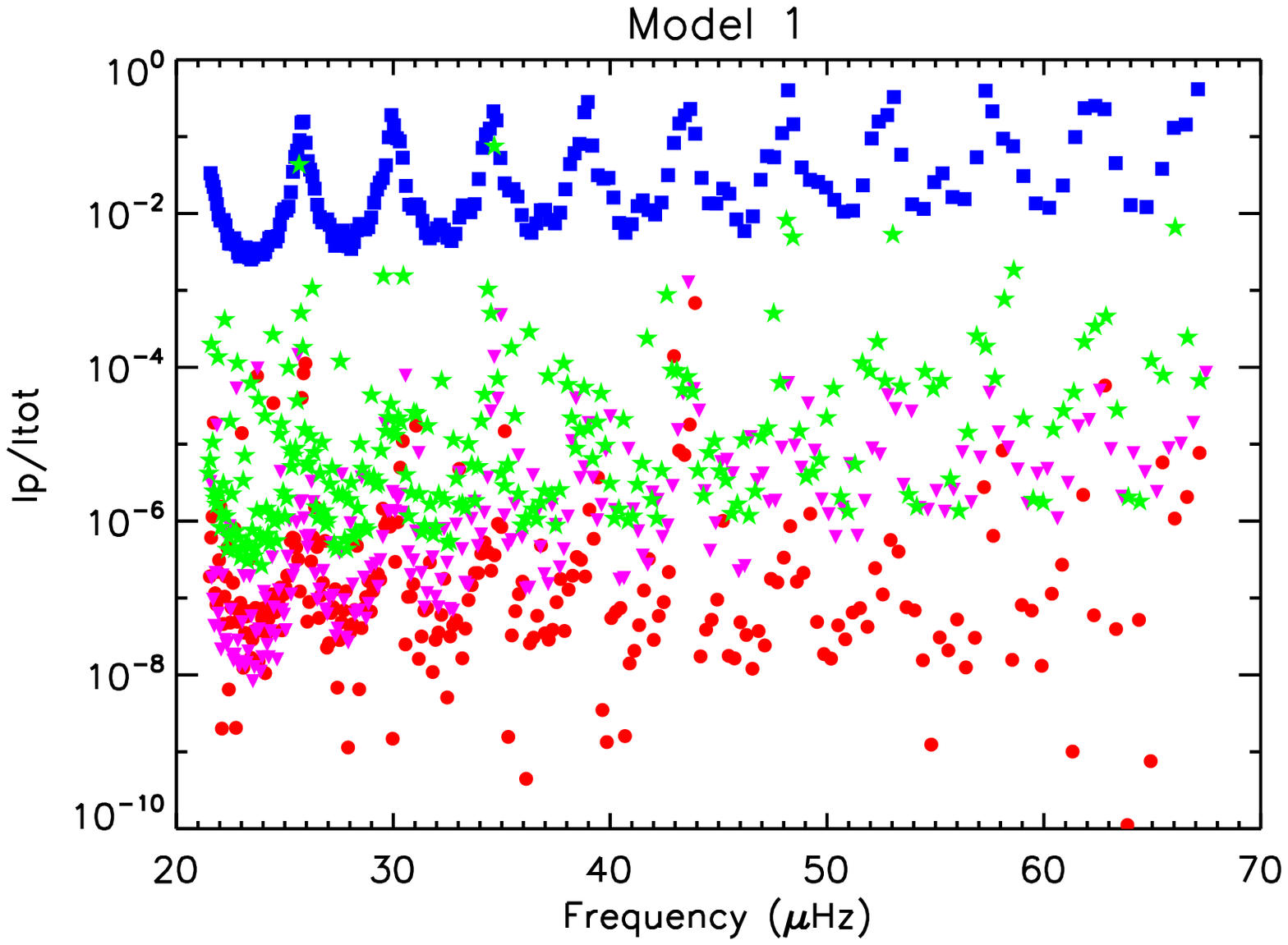}
\includegraphics[width=8.5cm]{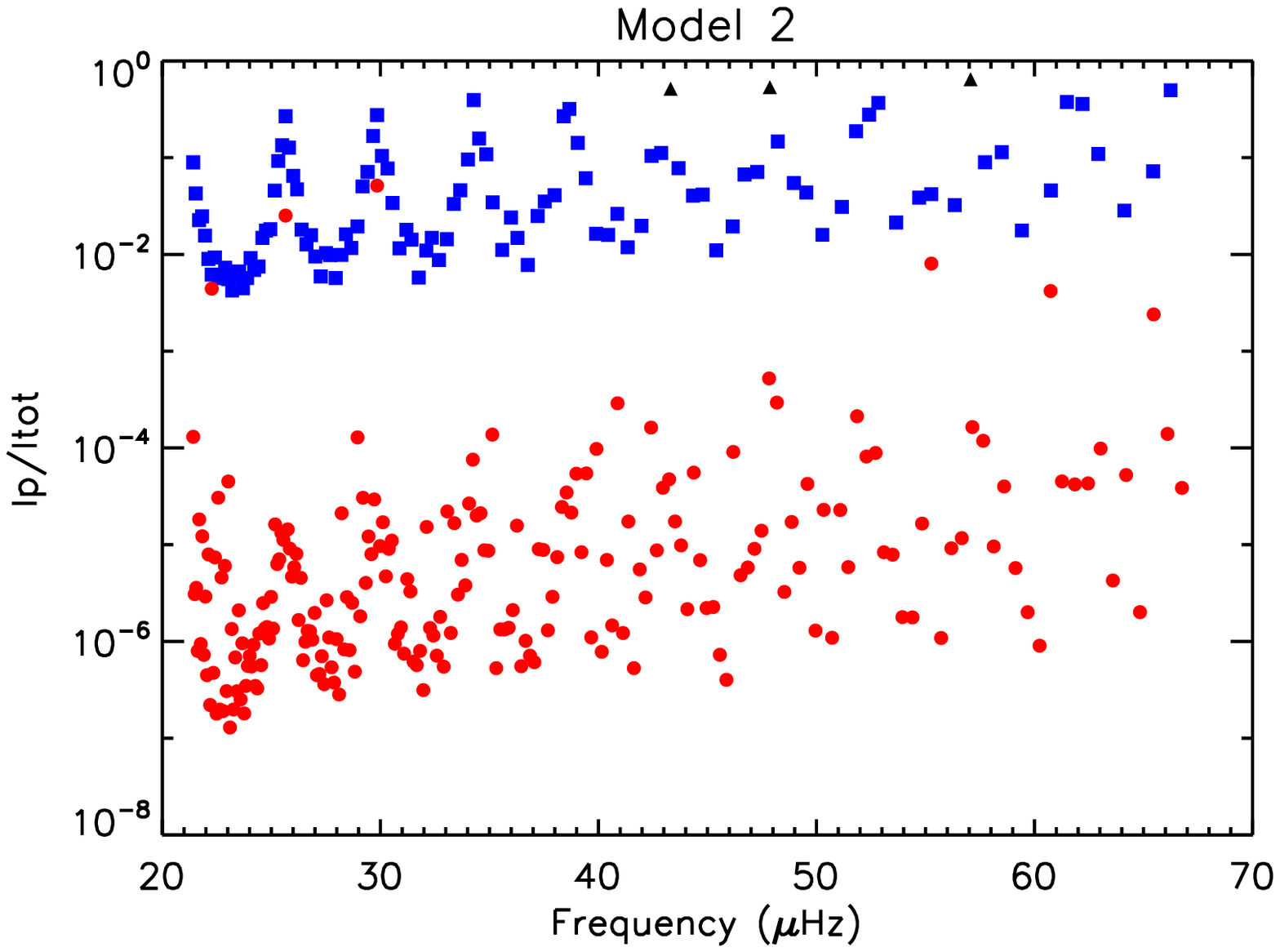}
\includegraphics[width=8.5cm]{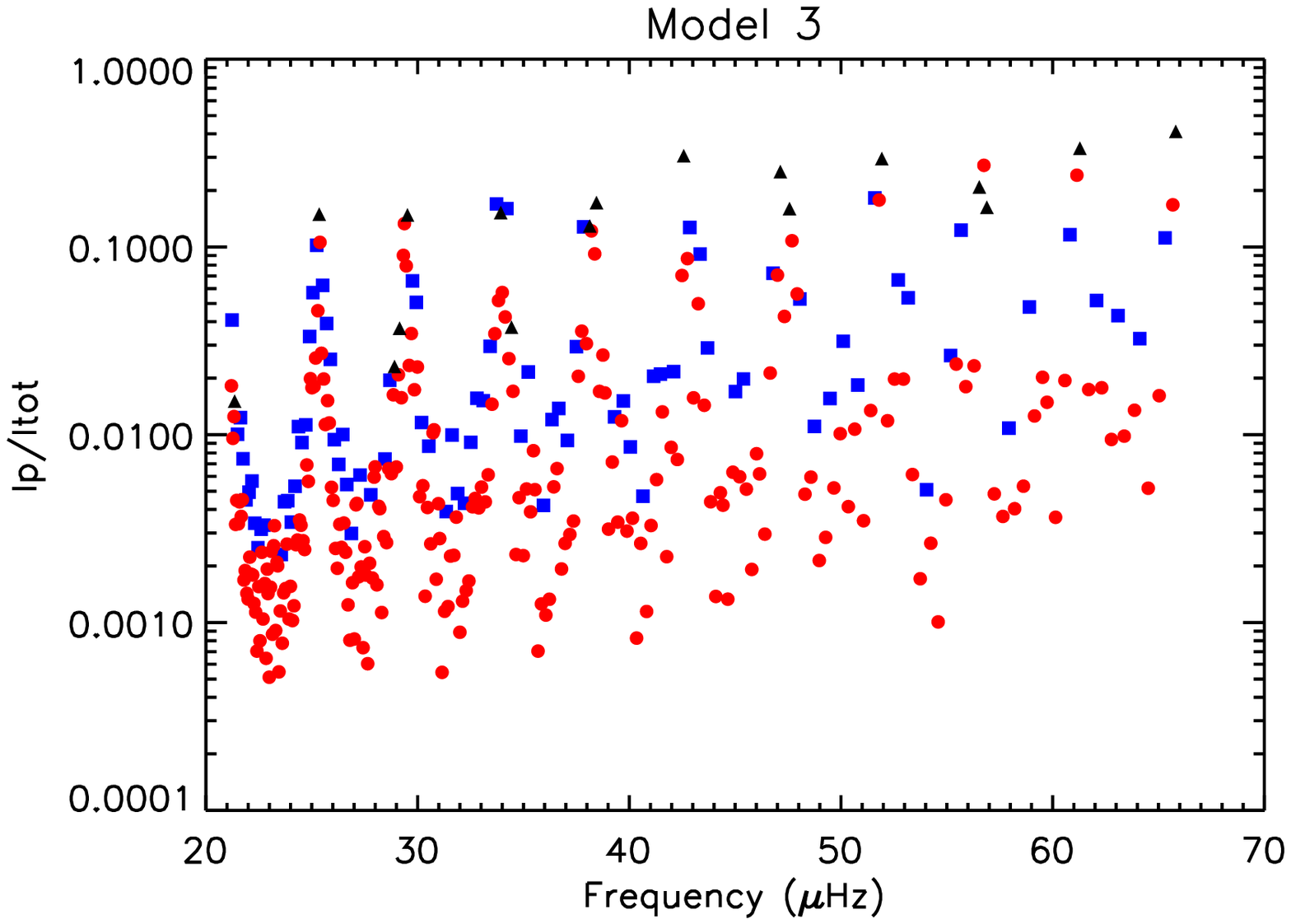}
\end{center}
\caption{Fractional contribution of the acoustic cavity to the mode inertia for models 1, 2, and 3. Modes that have more than half of their energy trapped in the \gun\ (resp. \gdeux) cavity are shown as filled red circles (resp. blue squares). Other modes are shown as filled black triangles. In the top panel, we also show $I_{\rm p}/I$ for the \gun-dominated modes of models 1th (purple downward triangles) and 1ov (green stars). \label{fig_ratio_adipls}}
\end{figure}

The ratios of inertia $I_{\rm p}/I$, giving the fractional contribution of the acoustic cavity to the total mode inertia, can be directly estimated using the mode eigenfunctions. They are shown for models 1,  1th, 1ov, 2, and 3 in Fig. \ref{fig_ratio_adipls}. For each model, the range of values of $I_{\rm p}/I$ taken by  \gun-dominated modes is indicative of the intensity of the coupling between the \gun-cavity and the rest of the star. The top panel of Fig. \ref{fig_ratio_adipls} confirms that this coupling is very weak for model 1. The coupling appears to be stronger for model 1th, in which the gradients of mean molecular weight at the boundaries of the He-burning shell are smoothed by thermohaline mixing. For model 1ov, in which the $\mu$-gradients are pushed away from the evanescent zone (see Sect. \ref{sect_coupling}), the coupling is several orders-of-magnitude stronger than in model 1. This confirms that the sharp $\mu$-gradients play an important role in the intensity of the coupling between the g-mode cavities.

The plots shown in Fig. \ref{fig_ratio_adipls} are qualitatively similar to those obtained with the JWKB approximation (Fig. \ref{fig_inertia_JWKB}). However, they confirm that the coupling strength between the two g-mode cavities is much weaker in numerical calculations than predicted by Eq. \ref{eq_q1} during most of the He subflash. For instance, for models 1ov and 2, Fig. \ref{fig_ratio_adipls} shows that the most \gun-like modes have a ratio $I_{\rm p}/I$ between $10^{-7}$ and $10^{-6}$. To reproduce this using the JWKB approximation, one needs $q_1\sim10^{-4}$, which is much weaker than the values of $q_1$ given by Eq. \ref{eq_q1} for these models ($q_1=0.10$ for model 1ov and $q_1=0.03$ for model 2). These discrepancies show that the asymptotic expression of $q_1$ given by Eq. \ref{eq_q1} is inappropriate. This is likely at least partly caused by the sharp $\mu$-gradients located near the edges of the evanescent zone, which produce non-negligible contributions of the derivatives of equilibrium quantities. The coupling strength would likely be better estimated by asymptotic calculations if the $\mu$-gradients were treated as glitch-like perturbations of the equilibrium quantities using a variational principle (e.g. \citealt{monteiro94}). This is however out of the scope of the present paper.


By contrast, for model 3, the \gun-dominated modes can reach large values of the ratio $I_{\rm p}/I$, which are comparable to those of \gdeux-dominated modes. This is consistent with a coupling strength $q_1=0.1$, as given by Eq. \ref{eq_q1}.

\subsection{Mode heights \label{sect_height_adipls}}

\begin{figure}
\begin{center}
\includegraphics[width=8.5cm]{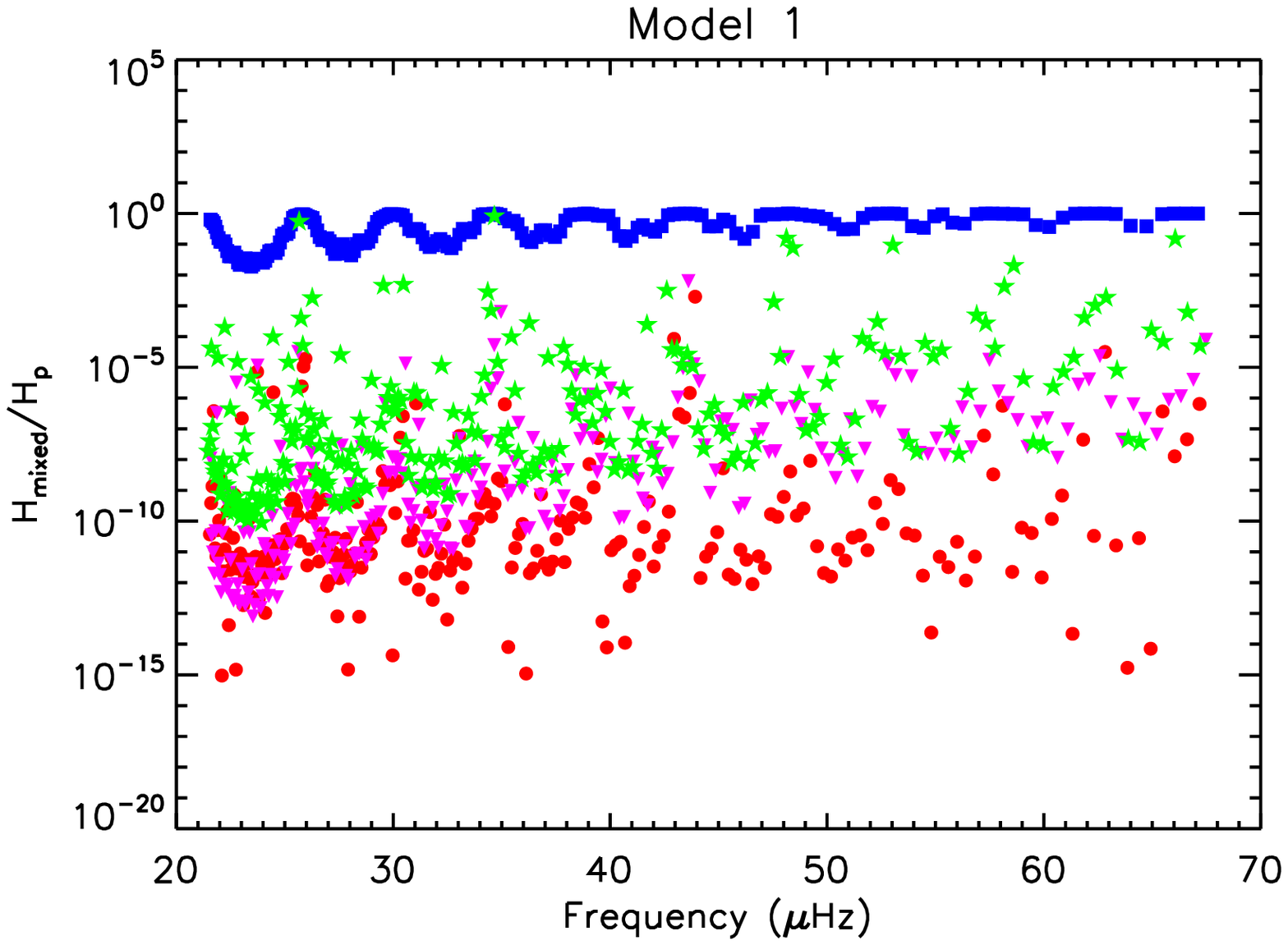}
\includegraphics[width=8.5cm]{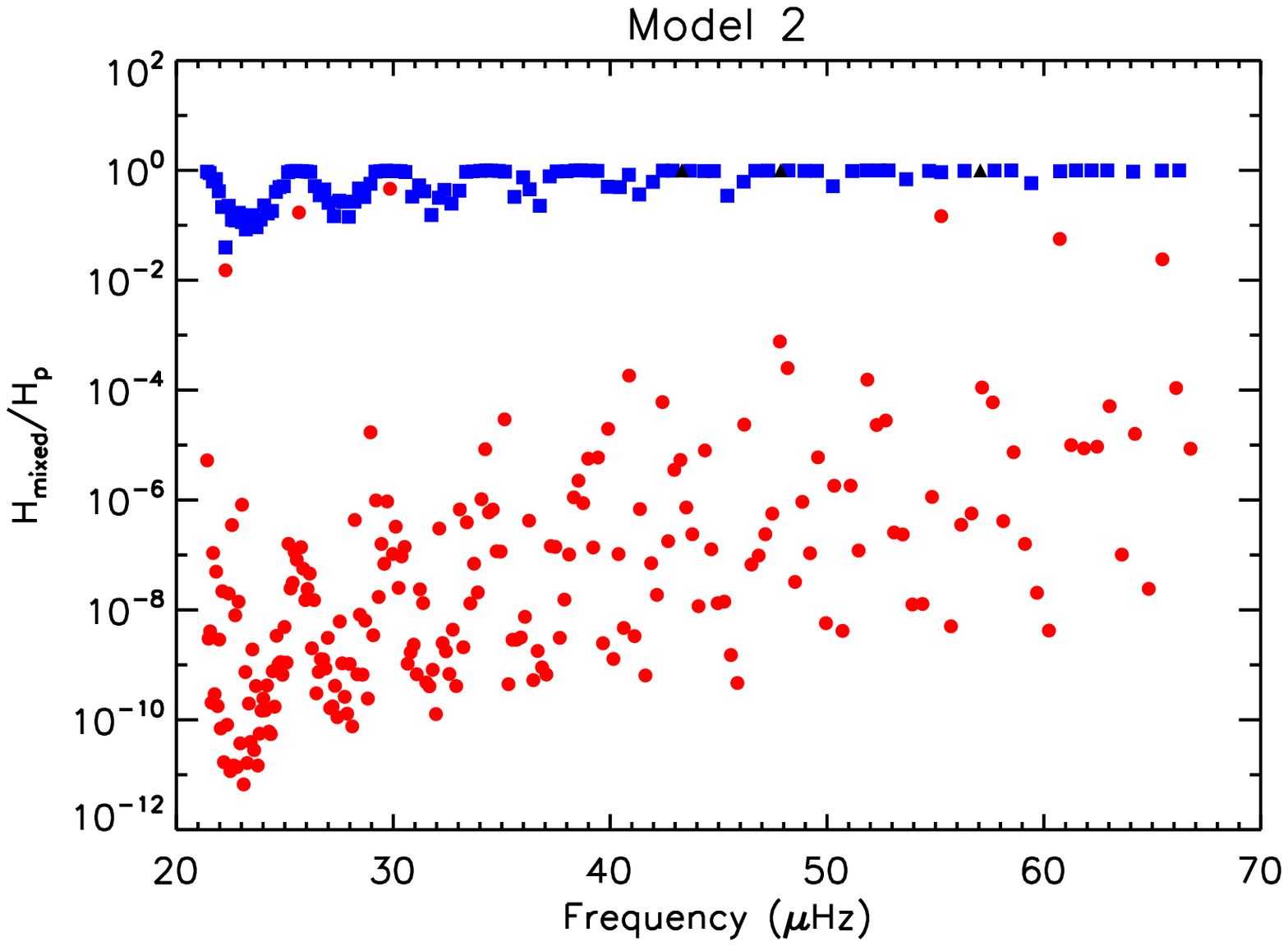}
\includegraphics[width=8.5cm]{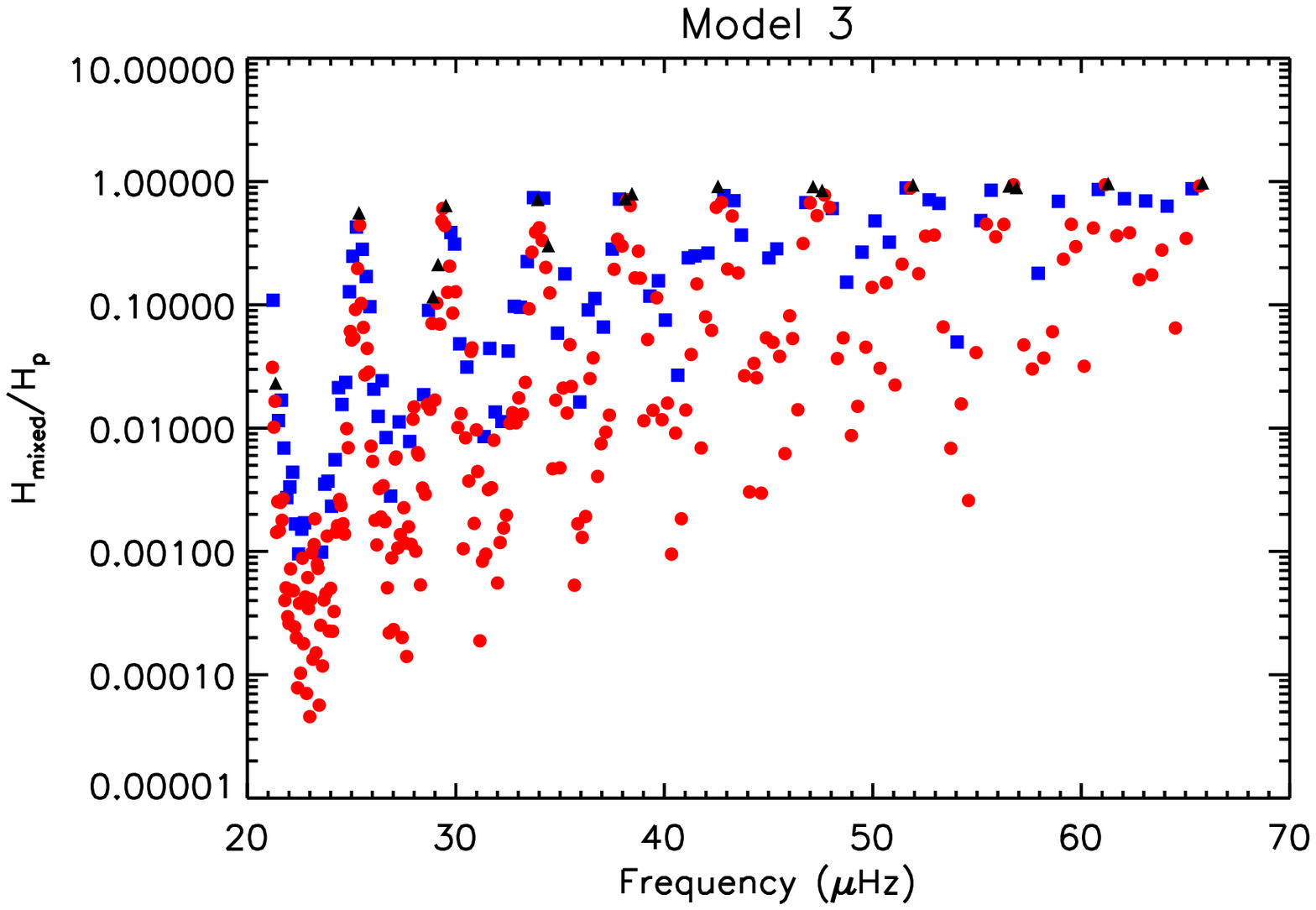}
\end{center}
\caption{Height ratios between $l=1$ mixed modes and theoretical $l=1$ pure p modes. for models 1, 1th, 1ov, 2, and 3. The symbols are the same as in Fig. \ref{fig_ratio_adipls}. \label{fig_height_adipls}}
\end{figure}

As was done in the framework of the JWKB approximation, we estimated the expected heights of the oscillation modes in the observed power spectrum based on our numerical calculations. For this purpose, we began by estimating the damping rates of the modes, taking into account non-adiabatic effects in the envelope and the contribution from radiative damping in the core. We used Eq. \ref{eq_etag}, in which the ratios of inertia $I_{\rm g1}/I$ and $I_{\rm g2}/I$ were obtained using the numerically computed mode eigenfunctions. The modes for which the damping rate exceeds the threshold $\eta_{\rm lim}=2/T_{\rm obs}$ are expected to be resolved.

To estimate the mode heights, we needed to calculate the ratio between the work integral in the g-mode cavities ($W_{\rm g1}+W_{\rm g2}$) and the work integral in the p-mode cavity $W_{\rm e}$. For this purpose, Eq. \ref{eq_work_ratio_JWKB} was rewritten as
\begin{equation}
\frac{W_{\rm g_1}+W_{\rm g_2}}{W_{\rm e}} = \frac{1}{4\pi\eta_{\rm p}} \left[ \frac{I_{\rm g1}}{I_{\rm p}}\frac{\mathcal{J}(\ra,\rb)}{\theta_{\rm g1}} + \frac{I_{\rm g2}}{I_{\rm p}}\frac{\mathcal{J}(\rc,\rd)}{\theta_{\rm g2}} \right],
\label{eq_work_ratio_adipls}
\end{equation}
where the ratios of inertia $I_{\rm g1}/I_{\rm p}$ and $I_{\rm g2}/I_{\rm p}$ were calculated using the mode eigenfunctions. We then used Eq. \ref{eq_res} and \ref{eq_unres} to estimate the heights of resolved and unresolved modes, respectively. The value of $\eta_{\rm p}$ obtained in Sect. \ref{sect_etap} was also used here.

Fig. \ref{fig_height_adipls} shows the ratio between the predicted heights of mixed modes and those of pure p modes for models 1, 1th, 1ov, 2, and 3. For model 1 and 1th, owing to the very weak coupling between the two g-mode cavities, all the modes that are trapped mostly in the \gun-cavity are expected to have negligible heights in the power spectrum 
On the contrary, for models 1ov, 2, and 3, we find that several \gun-dominated modes should have detectable heights in the power spectrum. For model 3, the coupling between the g-mode cavities is so large that all the modes have non-negligible contribution from the \gun-cavity ($I_{\rm g_1}/I > 0.2$ for all the modes). Consequently, we expect to have numerous \gun-dominated modes with detectable heights in the vicinity of pure p modes, as shown by the bottom right panel of Fig. \ref{fig_height}. Another consequence of the strong coupling between the two g-mode cavities in model 3 is that the \gdeux-dominated modes have a larger contribution from the \gun-cavity than in other models. This induces a decrease in their expected heights, compared to other models. As a result, in the case of a strong coupling between the g-mode cavities, we expect to detect modes almost exclusively in the neighborhood of pure p modes.


\begin{figure*}
\begin{center}
\includegraphics[width=\textwidth]{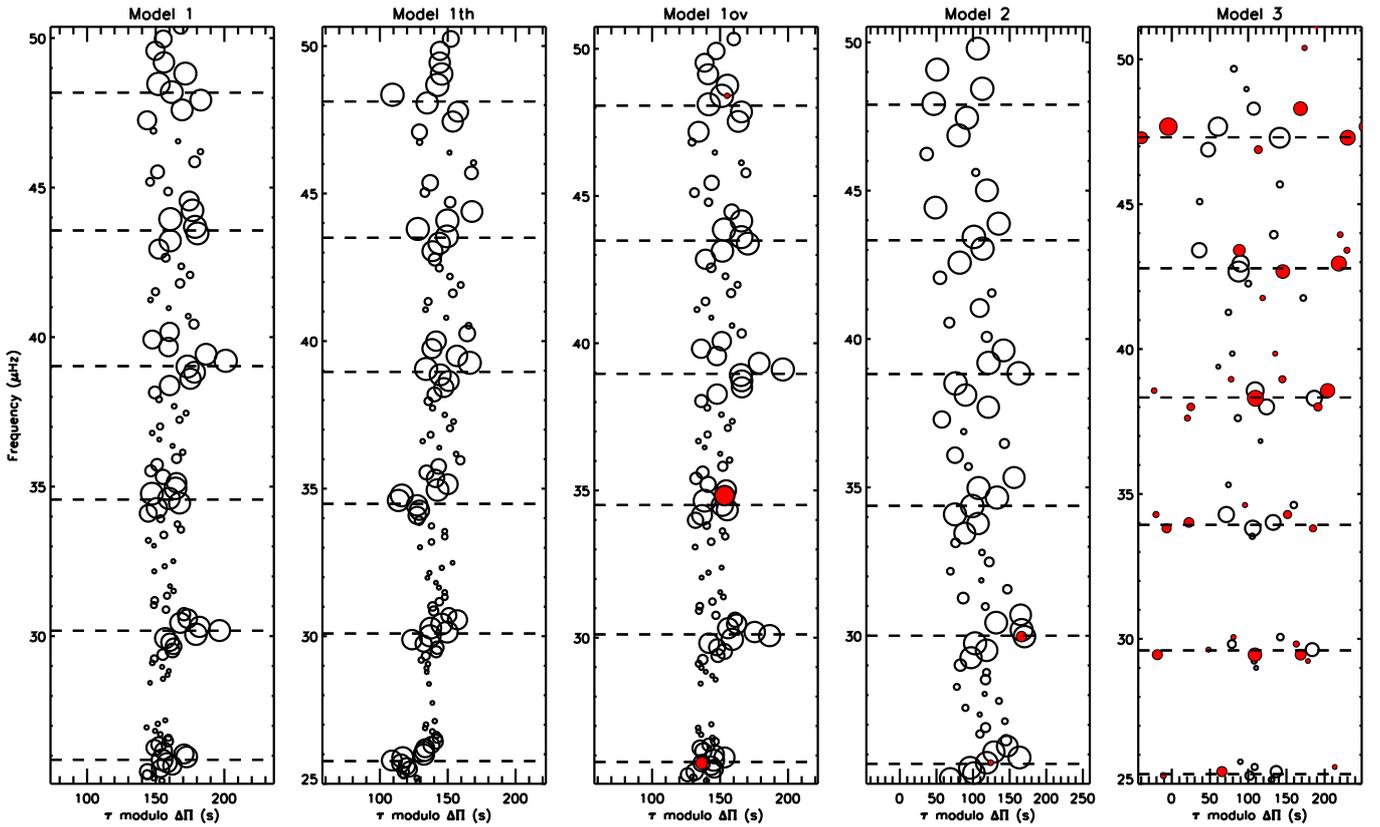}
\end{center}
\caption{Stretched \'echelle diagrams for models 1, 1th, 1ov, 2, and 3. The symbols have the same meaning as in Fig. \ref{fig_echelle_ratio}.
\label{echelle_stretch_adipls}}
\end{figure*}


\subsection{Stretched \'echelle diagrams}

As was already mentioned in Sect. \ref{sect_stretch_JWKB}, stretched period \'echelle diagrams are a very convenient representation of the oscillation spectrum in order to identify \gun-dominated mixed modes. To build such diagrams for each of the considered models, we used the \gdeux-dominated modes to estimate the $\zeta$ function, following the procedure described by \cite{mosser15}. We thus obtained stretched periods $\tau$ for all the modes, which we represented in an \'echelle diagram folded with $\Delta\Pi_2$. The stretched \'echelle diagrams for models 1, 1th, 1ov, 2, and 3 are shown in Fig. \ref{echelle_stretch_adipls}, where we plotted only the modes that have potentially detectable heights. As expected, most of the detectable modes regroup along a vertical ridge. This vertical ridge is not as straight as it was in our asymptotic calculations (see Fig. \ref{fig_echelle_ratio}). The oscillation that is clearly seen in the vertical ridge of models 1, 1ov, and 2 is caused by the buoyancy glitch produced by the H-burning shell in the \gdeux-cavity (see Sect. \ref{sect_glitch}).

Only at the end of the subflash (model 3) does the coupling between the g-mode cavities become large enough to produce additional detectable modes that significantly deviate from the asymptotic pattern of the \gdeux- and p-dominated modes, and thus lie clearly outside of the vertical ridge in the stretched \'echelle diagram.
During the other stages of the subflash, the coupling between the g-mode cavities is weaker, so that the stretched \'echelle diagram alone cannot yield direct detections of \gun-dominated modes. However, for models 1ov and 2, several modes that are mainly trapped in the \gun-cavity have detectable heights. We explain in the next section how this might be used to identify red giants going through a He subflash.

\section{Discussion and conclusion \label{sect_discussion}}

In this paper, we showed that red giants undergoing He core subflashes could be identified using the properties of their oscillation mode frequencies. During a He subflash, red giants have three propagating cavities because the convective He-burning shell splits the g-mode cavity in two. We calculated the expected mode frequencies in this case using both an asymptotic analysis and full numerical computations. We further estimated the expected mode heights taking into account the effects of radiative damping in order to determine which oscillation modes could be detected in \kepler\ seismic data.
 
If the \kepler\ sample does contain red giants that are undergoing a He subflash, these stars must have been so far identified as core-helium burning giants, considering the resemblance of their oscillation spectra. However, we have obtained in this study a list of clear, detectable features that could enable us to identify red giants passing through a He subflash. These features can be summarized as follows

\paragraph{Asymptotic period spacing between He subflashes}

\cite{bildsten12} showed that the asymptotic period spacings of g modes in the aftermath of a He subflash are significantly larger than those of RGB stars and much smaller than those of red clump stars (see also Fig. \ref{fig_deltapi_q}, top panel).

\paragraph{Asymptotic period spacing of the \gdeux-cavity during a He subflash}

We showed in Sect. \ref{sect_deltapi_JWKB} that during the first part of a He subflash, the asymptotic period spacing of the \gdeux-cavity also lies in the ``period spacing desert'' between RGB and red clump stars. As we have shown, most of the \gdeux-dominated modes should have detectable heights in the oscillation spectra of red giants going through a subflash, so that the asymptotic period spacing $\Delta\Pi_2$ should be easy to measure observationally. This could therefore constitute  a way of identifying red giants undergoing a He subflash. However, the shortness of the period during which $\Delta\Pi_2$ is significantly below the lowest period spacings of red clump stars makes this type of detection rather unlikely.

\paragraph{Buoyancy radius of the H-burning shell}

We have shown that the signature of the glitch produced by the H-burning shell in the \gdeux-cavity should be clearly detectable in the oscillation spectrum of red giants undergoing a He subflash (see Sect. \ref{sect_glitch}). It causes the periods of \gdeux-dominated modes to oscillate as a function of the radial order (see Fig. \ref{fig_echelle_adipls} and \ref{echelle_stretch_adipls}), and the period of this oscillation can be used to estimate the buoyancy radius of the H-burning shell in the \gdeux-cavity (see Eq. \ref{eq_period_osc}). During the He subflash, we found that the buoyancy radius of the H-burning shell corresponds to 82\% of the total buoyancy radius of the \gdeux-cavity, which results in an oscillation with a period of $\Delta n \sim 6$. By contrast, 
during the clump phase, the buoyancy radius of the H-burning shell amounts to about 70\% of the buoyancy radius of the g-mode cavity. This corresponds to an oscillation with a period of $\Delta n \sim 3$ for the g modes, which is significantly different from what was obtained during a He subflash. We thus conclude that the period of the oscillation produced by the H-burning shell could be used as a means to separate red giants undergoing a He subflash from regular clump stars.

\paragraph{Detection of additional modes trapped in the \gun-cavity}

Based on our study, we expect to detect mostly \gdeux- and p-dominated modes during a He subflash. Consequently, the oscillation spectrum of a red giant going through a He subflash should be quite similar to that of a regular clump star with an asymptotic period spacing corresponding to $\Delta\Pi_2$. However, during the most part of the subflash, several \gun-dominated mixed modes are also expected to have detectable heights in the power spectrum. These modes could appear as anomalies in the oscillation spectrum of red giants identified as belonging to the red clump. If the coupling $q_1$ between the g-mode cavities is large enough, the frequencies of the detectable \gun-dominated modes significantly deviate from the pattern of the \gdeux- and p-dominated modes. In this case, representing the oscillation modes in a stretched \'echelle diagram would enable us to directly spot these additional modes (see bottom right panel of Fig. \ref{fig_echelle_ratio} and right panel of Fig. \ref{echelle_stretch_adipls}). This seems to be an efficient way of identifying stars in a He subflash. However, we found that $q_1$ becomes strong enough to make this type of detection possible only at the end of a He subflash. The probability of detecting a red giant in this specific phase remains low. We should also mention the caveat that additional modes can also be produced by structural glitches in certain conditions (\citealt{cunha15}). Although we did not find such effects in the stellar models that were computed in this study, this should be kept in mind when searching for red giants going through the He flash using seismology.

\paragraph{Detection of anomalous rotational splittings}

In the case of weaker coupling intensities $q_1$, the detectable \gun-dominated modes are not expected to show significant deviations from the pattern of the \gdeux- and p-dominated modes (see middle panels of Fig. \ref{echelle_stretch_adipls}). However, for several of the detectable mixed modes, the \gun-cavity contributes to more than half of the inertia. One important consequence is that these modes are predominantly sensitive to the rotation in the \gun-cavity. Predicting the rotation profile of red giants during the He-core flash would require to determine (i) the internal rotation profile of red giants at the tip of the RGB (which is unknown because seismology has brought measurements of the rotation of RGB stars no further than the luminosity bump), (ii) the structural changes during the He-flash (expansion of the layers below the H-burning shell by about a factor of 10 and contraction of the layers above), and (iii) the internal transport of angular momentum during this phase, which is completely unknown. The structural changes during the flash induce a strong forcing of radial differential rotation between the layers below the H-burning shell and those above. Unless the redistribution of angular momentum is efficient enough to enforce a solid-body rotation during the flash despite the shortness of this phase, one expects to have a different rotation rate below and above the H-burning shell. If it is indeed the case, then the average rotation in the \gdeux-mode cavity (which includes the H-burning shell) is expected to be different from the rotation in the \gun-cavity, which lies well below the H-burning shell. It is thus plausible that \gun-dominated modes have rotational splittings that differ from those of the \gdeux-dominated modes. This could be used to identify \gun-dominated modes in the oscillation spectra.

\begin{figure}
\begin{center}
\includegraphics[width=9cm]{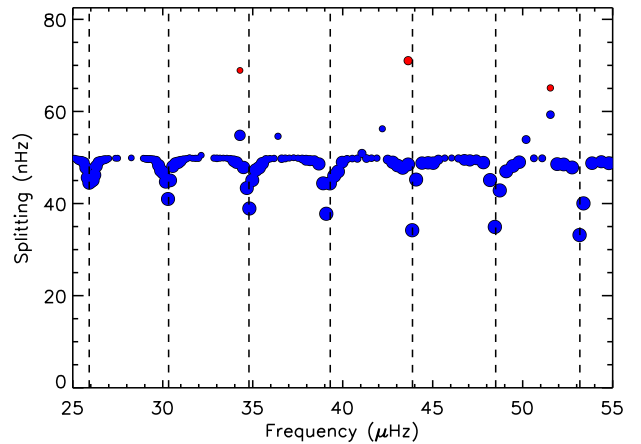}
\end{center}
\caption{Rotational splittings obtained for model 1ov, assuming a piecewise-constant rotation profile with $\Omega_{\rm g_1} = 150$ nHz, $\Omega_{\rm g_2} = 100$ nHz, and $\Omega_{\rm p} = 10$ nHz (see text). Modes that are trapped mainly in the \gun-cavity (resp. \gdeux-cavity) are shown as red (resp. blue) filled circles. The size of the circles indicates the expected height of each mode in the observed spectrum. Only the modes with a height corresponding to at least 10\% of the height of a pure p mode are shown. The vertical dashed lines indicate the location of theoretical pure p modes.
\label{fig_splitting}}
\end{figure}

As mentioned above, the internal rotation profiles of red giants during the He flash are completely unknown. We only know from \cite{mosser12b} that the core of primary clump stars spins with an average frequency of about 100 nHz. In order to roughly assess the sensitivity of the diagnostic that we propose here, we considered a piecewise-constant rotation profile with the \gdeux-cavity spinning at a rotation rate $\Omega_{\rm g_2} = 100$ nHz, the p-mode cavity rotating at $\Omega_{\rm p} = 10$ nHz (this value is completely arbitrary, but it does not impact our conclusions), and the \gun-cavity rotating only 50\% faster than the \gdeux-cavity, i.e. $\Omega_{\rm g_1} = 150$ nHz. We calculated the rotational kernels of model 1ov and combined them with the chosen rotation profile to calculate theoretical rotational splittings, which are shown in Fig. \ref{fig_splitting}. As can be seen in Fig. \ref{fig_splitting}, the splittings of \gdeux- and p-dominated modes vary smoothly with frequency\footnote{In the case of two mode cavities, it has been shown that the variations in the rotational splittings with frequency correspond to the so-called $\zeta$ function (see \citealt{mosser15} for more details).}, while modes that have a non-negligible contribution of the \gun-cavity to their inertia have rotational splittings that clearly deviate from this relation. With the chosen rotation profile, the splittings of pure \gun\ modes would be approximately $\Omega_{\rm g_1}/2 = 75$ nHz (see e.g. \citealt{goupil13}), a value which is nearly reached by the most \gun-dominated modes (see Fig. \ref{fig_splitting}). This represents a 25-nHz difference with the splittings of \gdeux-dominated modes. With four years of \kepler\ data, the expected precision on the measured splittings is about 10 nHz, so that \gun-dominated would have significantly larger rotational splittings in the observed spectrum. Red giants going through a He subflash could thus be characterized by the detection of mixed dipolar modes with rotational splittings that are significantly different from those of the \gdeux- and p-dominated modes, which follow a well-understood pattern (\citealt{mosser15}). \\

These features can already be searched for within the catalog of about 15,000 \kepler\ red giants for which oscillations have been detected. The seismic detection of He flashing giants would nicely confirm the existence of the He-core subflashes predicted by 1D evolutionary models and exclude the picture based on 2D- and 3D-simulations claiming no occurence of a series of subflashes (\citealt{mocak08}, \citealt{mocak09}). If rotational splittings can be detected for mixed modes trapped mainly in the inner g-mode cavity during a He-subflash, we could also measure the differential rotation within the core of the star. This would place precious constraints on the transport of angular momentum in red giants. Finally, we stress that the detection of the seismic features of the He subflashes depends a lot on the intensity of the coupling between the two g-mode cavities, which remains uncertain. Consequently, even if these features fail to be detected with \kepler\ data, it would not necessarily exclude the picture of the He flash based on 1D evolutionary models.

\begin{acknowledgements}
We wish to thank the referee, H. Shibahashi, for comments and suggestions that significantly improved the clarity of the manuscript. We are very thankful to B. Mosser, J. Christensen-Dalsgaard, M.~J. Goupil, and M. Vrard for fruitful discussions about the paper. S.D. acknowledges support from the PNPS under the grant ``Rotation interne et magn\'etisme des sous-g\'eantes et g\'eantes \kepler'' and from the Centre National d'Etudes Spatiales (CNES).
\end{acknowledgements}

\bibliographystyle{aa.bst} 
\bibliography{biblio} 

\onecolumn

\begin{appendix}

\section{Asymptotic analysis for three propagation cavities \label{app_JWKB}}

Eq. \ref{eq_v} and \ref{eq_w} are solved around each turning point, and the solutions are then matched at an intermediate radius between each pair of turning points. We first write the expressions of $v$ and $w$ in the vicinity of the inner g-mode cavity (\gun), which extends between $\ra$ $\rb$. Since $\omega=N$ in $\ra$ and $\rb$, Eq. \ref{eq_w} is singular at these radii, and we thus solve Eq. \ref{eq_v}. The solutions are given by:
\begin{numcases}{v \sim}
  	& $\displaystyle \frac{a}{2\sqrt{\pi\kappa}} \exp \left( -\int_r^{\ra} \kappa \dr \right) + \frac{b}{\sqrt{\pi\kappa}} \exp \left( \int_r^{\ra} \kappa \dr \right), \;\;\; \hbox{for}\; r \ll \ra$ \\
  	& $\displaystyle \frac{a}{\sqrt{\pi k_r}} \cos \left( \int_{\ra}^r k_r \dr - \frac{\pi}{4} \right) - \frac{b}{\sqrt{\pi k_r}} \sin \left( \int_{\ra}^r k_r \dr - \frac{\pi}{4} \right), \;\;\; \hbox{for}\; r \gg \ra$ \label{eq_vg1} \\ 
	& $\displaystyle \frac{c}{\sqrt{\pi k_r}} \cos \left( \int_r^{\rb} k_r \dr - \frac{\pi}{4} \right) -
	\displaystyle \frac{d}{\sqrt{\pi k_r}} \sin \left( \int_r^{\rb} k_r \dr - \frac{\pi}{4} \right), \;\;\; \hbox{for}\; r \ll \rb$ \\
	& $\displaystyle \frac{c}{2\sqrt{\pi\kappa}} \exp \left( -\int_{\rb}^r \kappa \dr \right) +
	\displaystyle \frac{d}{\sqrt{\pi\kappa}} \exp \left( \int_{\rb}^r \kappa \dr \right) , \;\;\; \hbox{for}\; r \gg \rb$
\end{numcases}
To ensure the regularity of $v$ in the center, the coefficient $b$ must be equal to zero. The expression of $w$ in the vicinity of the \gun\ cavity can be obtained from the relation (cf \citealt{shibahashi79})
\begin{equation}
w=\frac{1}{|k_r|} \frac{S_l^2-\omega^2}{|S_l^2-\omega^2|}\deriv{v}{r},
\label{eq_wfromv}
\end{equation}
which yields
\begin{numcases}{w \sim}
  	& $\displaystyle \frac{a}{2\sqrt{\pi\kappa}} \exp \left( -\int_r^{\ra} \kappa \dr \right), \;\;\; \hbox{for}\; r \ll \ra$ \\
  	& $\displaystyle -\frac{a}{\sqrt{\pi k_r}} \sin \left( \int_{\ra}^r k_r \dr - \frac{\pi}{4} \right), \;\;\; \hbox{for}\; r \gg \ra$ \\
	& $\displaystyle \frac{c}{\sqrt{\pi k_r}} \sin \left( \int_r^{\rb} k_r \dr - \frac{\pi}{4} \right) + 
	\displaystyle \frac{d}{\sqrt{\pi k_r}} \cos \left( \int_r^{\rb} k_r \dr - \frac{\pi}{4} \right), \;\;\; \hbox{for}\; r \ll \rb$ \\
	& $\displaystyle -\frac{c}{2\sqrt{\pi\kappa}} \exp \left( -\int_{\rb}^r \kappa \dr \right) + 
	\displaystyle \frac{d}{\sqrt{\pi\kappa}} \exp \left( \int_{\rb}^r \kappa \dr \right) , \;\;\; \hbox{for}\; r \gg \rb$
\end{numcases}
The matching of $v$ and $w$ inside the first g-mode cavity imposes the relations
\begin{align}
a \cos \left( \int_{\ra}^r k_r \dr - \frac{\pi}{4} \right) & = c \cos \left( \int_r^{\rb} k_r \dr - \frac{\pi}{4} \right) - d \sin \left( \int_r^{\rb} k_r \dr - \frac{\pi}{4} \right) \\
- a \sin \left( \int_{\ra}^r k_r \dr - \frac{\pi}{4} \right) & = c \sin \left( \int_r^{\rb} k_r \dr - \frac{\pi}{4} \right) - d \cos \left( \int_r^{\rb} k_r \dr - \frac{\pi}{4} \right)
\end{align}
which yields
\begin{align}
c & = a \sin \tgun \\
d & = a \cos \tgun
\end{align}
where $\tgun \equiv \int_{\ra}^{\rb} k_r \dr$. In the case of a single g-mode cavity, one would need to impose $d=0$ for regularity at the surface, and we would thus obtain the eigenvalue condition $\cos \tgun = 0$, which corresponds to pure g modes. \\

The expressions of the eigenfunctions in the shallower g-mode cavity (\gdeux) are similar to those of the \gun-cavity. We introduce the coefficients $(a',b',c',d')$ analogous to the coefficients $(a,b,c,d)$ corresponding to the \gun-cavity. The only difference with the \gun-cavity is that the coefficient $b'$ of the term $\exp \left( \int_r^{\rc} \kappa \dr \right)$ in the evanescent zone does not vanish.
The matching of $v$ and $w$ in the \gdeux-cavity eventually requires that:
\begin{align}
c' & = a' \sin \tgdeux + b' \cos \tgdeux \\
d' & = a' \cos \tgdeux - b' \sin \tgdeux 
\end{align}
where we have introduced $\tgdeux \equiv \int_{\rc}^{\rd} k_r \dr$. \\

In the p-mode cavity, we first solve Eq. \ref{eq_w} for $w$ around $\re$ and we obtain $v$ from the relation
\begin{equation}
v=\frac{1}{|k_r|} \frac{\omega^2-N^2}{ |\omega^2-N^2|} \deriv{w}{r}.
\end{equation}
We then solve Eq. \ref{eq_v} for $v$ around $\rf$ and deduce $w$ from Eq. \ref{eq_wfromv}. Introducing coefficients $(a'',b'',c'',d'')$ analogous to the coefficients of the g-mode cavities, we obtain
\begin{numcases}{v \sim}
  	& $\displaystyle \frac{a''}{2\sqrt{\pi\kappa}} \exp \left( -\int_r^{\re} \kappa \dr \right) - 
	\displaystyle \frac{b''}{\sqrt{\pi\kappa}} \exp \left( \int_r^{\re} \kappa \dr \right), \;\;\; \hbox{for}\; r \ll \re$ \\
  	& $\displaystyle -\frac{a''}{\sqrt{\pi k_r}} \sin \left( \int_{\re}^r k_r \dr - \frac{\pi}{4} \right) -
	\displaystyle \frac{b''}{\sqrt{\pi k_r}} \cos \left( \int_{\re}^r k_r \dr - \frac{\pi}{4} \right), \;\;\; \hbox{for}\; r \gg \re$ \\ 
	& $\displaystyle \frac{c''}{\sqrt{\pi k_r}} \cos \left( \int_r^{\rf} k_r \dr - \frac{\pi}{4} \right), \;\;\; \hbox{for}\; r \ll \rf$ \label{eq_v_p} \\
	& $\displaystyle \frac{c''}{2\sqrt{\pi\kappa}} \exp \left( -\int_{\rf}^r \kappa \dr \right), \;\;\; \hbox{for}\; r \gg \rf$
\end{numcases}
\begin{numcases}{w \sim}  
	& $\displaystyle \frac{a''}{2\sqrt{\pi\kappa}} \exp \left( -\int_r^{\re} \kappa \dr \right) +
	\displaystyle \frac{b''}{\sqrt{\pi\kappa}} \exp \left( \int_r^{\re} \kappa \dr \right), \;\;\; \hbox{for}\; r \ll \re$ \\
  	& $\displaystyle \frac{a''}{\sqrt{\pi k_r}} \cos \left( \int_{\re}^r k_r \dr - \frac{\pi}{4} \right) -
	\displaystyle \frac{b''}{\sqrt{\pi k_r}} \sin \left( \int_{\re}^r k_r \dr - \frac{\pi}{4} \right), \;\;\; \hbox{for}\; r \gg \re$ \\ 
	& $\displaystyle -\frac{c''}{\sqrt{\pi k_r}} \sin \left( \int_r^{\rf} k_r \dr - \frac{\pi}{4} \right), \;\;\; \hbox{for}\; r \ll \rf$ \\
	& $\displaystyle \frac{c''}{2\sqrt{\pi\kappa}} \exp \left( -\int_{\rf}^r \kappa \dr \right), \;\;\; \hbox{for}\; r \gg \rf$
\end{numcases}
The matching of $v$ and $w$ in the p-mode cavity imposes that:
\begin{align}
c'' \cos \left( \int_r^{\rf} k_r \dr - \frac{\pi}{4} \right) & = - a'' \sin \left( \int_{\re}^r k_r \dr - \frac{\pi}{4} \right) - b'' \cos \left( \int_{\re}^r k_r \dr - \frac{\pi}{4} \right) \\
- c'' \sin \left( \int_r^{\rf} k_r \dr - \frac{\pi}{4} \right) & = a'' \cos \left( \int_{\re}^r k_r \dr - \frac{\pi}{4} \right) - b'' \sin \left( \int_{\re}^r k_r \dr - \frac{\pi}{4} \right)
\end{align}
which eventually yields
\begin{align}
a'' & = c'' \cos\tetap \\
b'' & = - c'' \sin\tetap \nonumber
\end{align}
where $\tetap\equiv\int_{\re}^{\rf} k_r \dr$. \\

We then join the three regions by imposing that $v$ and $w$ match in the evanescent zones $[\rb,\rc]$ and $[\rd,\re]$. This gives four more relations
\begin{align}
\frac{a'}{2} \exp\left( -\int_r^{\rc} \kappa \dr \right) + b' \exp\left( \int_r^{\rc} \kappa \dr \right) & = \frac{a}{2} \sin\tgun \exp\left( -\int_{\rb}^r \kappa \dr \right) + a \cos\tgun \exp\left( \int_{\rb}^r \kappa \dr \right) \label{eq_syst_app1} \\
\frac{a'}{2} \exp\left( -\int_r^{\rc} \kappa \dr \right) - b' \exp\left( \int_r^{\rc} \kappa \dr \right) & = - \frac{a}{2} \sin\tgun \exp\left( -\int_{\rb}^r \kappa \dr \right) + a \cos\tgun \exp\left( \int_{\rb}^r \kappa \dr \right) \\
\frac{c'}{2} \exp\left( -\int_{\rd}^r \kappa \dr \right) + d' \exp\left( \int_{\rd}^r \kappa \dr \right) & = \frac{c''}{2} \cos\tetap \exp\left( -\int_r^{\re} \kappa \dr \right) + c'' \sin\tetap \exp\left( \int_r^{\re} \kappa \dr \right) \\
- \frac{c'}{2} \exp\left( -\int_{\rd}^r \kappa \dr \right) + d' \exp\left( \int_{\rd}^r \kappa \dr \right) & = \frac{c''}{2} \cos\tetap \exp\left( -\int_r^{\re} \kappa \dr \right) - c'' \sin\tetap \exp\left( \int_r^{\re} \kappa \dr \right)  \label{eq_syst_app4}
\end{align}
which eventually yields the eigenvalue relation:
\begin{equation}
\cot\tgun \cot\tgdeux \tan\tetap - q_2\cot\tgun - q_1\tan\tetap - q_1 q_2 \cot\tgdeux = 0 \label{eq_3cav_app}
\end{equation}
where 
$q_1 \equiv \frac{1}{4} \exp \left( -2 \int_{\rb}^{\rc} \kappa \dr \right)$ and 
$q_2 \equiv \frac{1}{4} \exp \left( -2 \int_{\rd}^{\re} \kappa \dr \right)$.
We also obtain the following relations between $a$, $c'$, $d'$ and $c''$
\begin{align}
c' & = \frac{c'' \sin\tetap}{\sqrt{q_2} } \label{eq_c1_c2} \\
d' & = c'' \sqrt{q_2} \cos\tetap  \label{eq_d1_c2} \\
\displaystyle a & = c'' \frac{\sin\tetap\cos\tgdeux-q_2\cos\tetap\sin\tgdeux}{\sin\tgun\sqrt{q_1 q_2}}.
\label{eq_a_c2}
\end{align}


\section{Ratios of inertia \label{app_ratios}}

Analogously to the approach of \cite{goupil13}, we seek to estimate the ratios $\alpha_1 \equiv I_{\rm g_1}/I_{\rm p}$ and $\alpha_2 \equiv I_{\rm g_2}/I_{\rm p}$. We follow the same procedure as these authors, and refer the reader to their paper for more details. 

We first estimate the inertia in the p-mode cavity by making the assumption that the radial motion is dominant over the horizontal motion in this region, so that
\begin{equation}
I_{\rm p} \sim 4\pi \int_{\re}^{\rf} \frac{v^2}{c_{\rm s}^2} \left| 1-\frac{S_l^2}{\omega^2} \right| \,\dr
\label{eq_Ip}
\end{equation}
The second equality has been obtained using Eq. \ref{eq_v}. We then use the expression of $v$ in the p-mode cavity given by Eq. \ref{eq_v_p}, where we define $\theta\equiv\int_r^{\rf} k_r \,\dr$. Assuming that $k_r\sim(\omega^2-S_l^2)^{1/2}/c_{\rm s}$ in the p-mode cavity and changing the integration variable to $\theta$ in Eq. \ref{eq_Ip}, we obtain
\begin{align}
I_{\rm p} & \sim \frac{2 c''^2}{\omega^2} \int_0^{\tetap} \left[ 1 + \sin(2\theta) \right] \,\hbox{d}\theta \\
&  \sim \frac{2 c''^2}{\omega^2} \left[ \tetap + \frac{1-\cos(2\tetap)}{2} \right] \approx \frac{2 c''^2\tetap}{\omega^2} \label{eq_Ip2}
\end{align}
In the last equality, we have neglected factors of order unity compared to $\tetap$, which is valid because we consider modes with large radial order $n$. We recover the expression obtained by \cite{goupil13} for the inertia in the p-mode cavity.

In the external g-mode cavity (delimited by $\rc$ and $\rd$), the inertia can be estimated by assuming that the horizontal motion is dominant over the radial motion, so that
\begin{equation}
I_{\rm g_2} \sim 4\pi \int_{\rc}^{\rd} l(l+1) \frac{\left| N^2-\omega^2 \right|}{r^2 \omega^4} w^2 \,\dr
\end{equation}
where we have used Eq. \ref{def_w}. We use the expression of $w$ in the $g_2$ cavity given in Appendix \ref{app_JWKB} and we define $\theta\equiv\int_r^{\rd} k_r \,\dr$. By assuming that 
\begin{equation}
k_r\sim \frac{\sqrt{l(l+1)}\left| N^2-\omega^2 \right|^{1/2}}{\omega r}
\end{equation}
in this g-propagative region, and using $\theta$ as an integration variable, we obtain
\begin{equation}
I_{\rm g_2} \sim \frac{2}{\omega^2} \left[ (c'^2+d'^2) \tgdeux + (c'^2-d'^2)\frac{\cos(2\tgdeux)-1}{2} -c'd'\sin(2\tgdeux) \right] \approx \frac{2 (c'^2+d'^2)\tgdeux}{\omega^2}
\label{eq_Ig2}
\end{equation}
As was done in the p-mode cavity, we have assumed that $\tgdeux\gg1$.

We proceed similarly to estimate the inertia in the inner g-mode cavity (delimited by $\ra$ and $\rb$), and obtain
\begin{equation}
I_{\rm g_1} \sim \frac{2 a^2}{\omega^2} \left[ \tgun + \frac{\cos(2\tgun)-1}{2} \right] \approx \frac{2 a^2\tgun}{\omega^2}
\label{eq_Ig1}
\end{equation}

The ratios of inertias $\alpha_1$ and $\alpha_2$ are thus given by
\begin{align}
\alpha_2 & = \frac{c'^2+d'^2}{c''^2}\frac{\tgdeux}{\tetap} = \frac{\sin^2\tetap + q_2^2\cos^2\tetap}{q_2} \frac{\Delta\nu}{\nu^2\Delta\Pi_2} \\
\alpha_1 & = \frac{a^2}{c''^2}\frac{\tgun}{\tetap} = \frac{\left( \sin\tetap\cos\tgdeux - q_2\cos\tetap\sin\tgdeux \right)^2}{q_1 q_2 \sin^2\tgun} \frac{\Delta\nu}{\nu^2\Delta\Pi_1} \label{eq_alpha1}
\end{align}
We have used the ratios between the amplitude of the eigenfunctions as obtained from JWKB analysis (Eq. \ref{eq_c1_c2}, \ref{eq_d1_c2}, and \ref{eq_a_c2}). We have also used the approximate expressions $\tetap\sim\pi\nu/\Delta\nu$ and $\theta_{{\rm g}_i}\sim\pi/(\nu\Delta\Pi_i)$ ($i=1,\,2$),  as advocated by \cite{goupil13}. We note that these approximations are not valid when these quantities are in the phase of a sine function. 

Using these expressions, we were able to calculate the fraction of the total energy of the mode that is trapped in each of the three cavities:
\begin{align}
\frac{I_{\rm p}}{I} & = \frac{1}{1+\alpha_1+\alpha_2} \label{eq_IpvsI} \\
\frac{I_{\rm g_1}}{I} & = \frac{\alpha_1}{1+\alpha_1+\alpha_2} \label{eq_Ig1vsI} \\
\frac{I_{\rm g_2}}{I} & = \frac{\alpha_2}{1+\alpha_1+\alpha_2} \label{eq_Ig2vsI} 
\end{align}

Note that the expression of $\alpha_1$ given by Eq. \ref{eq_alpha1} becomes singular when $\sin\tgun$ vanishes. Whenever this is the case, one can always rewrite Eq. \ref{eq_IpvsI} through \ref{eq_Ig2vsI} as
\begin{align}
\frac{I_{\rm p}}{I} & = \frac{\beta_1}{1+\beta_1+\beta_2} \label{eq_IpvsI_alt} \\
\frac{I_{\rm g_1}}{I} & = \frac{1}{1+\beta_1+\beta_2} \label{eq_Ig1vsI_alt} \\
\frac{I_{\rm g_2}}{I} & = \frac{\beta_2}{1+\beta_1+\beta_2} \label{eq_Ig2vsI_alt} 
\end{align}
where $\beta_1 \equiv I_{\rm p}/I_{\rm g_1}$ and $\beta_2 \equiv I_{\rm g_2}/I_{\rm g_1}$. We have $\beta_1=1/\alpha_1$, which is no longer singular, and using the expressions from Appendix \ref{app_JWKB}, one obtains:
\begin{equation}
\beta_2  = \frac{c'^2+d'^2}{a^2}\frac{\tgdeux}{\tgun} = \frac{\cos^2\tgun + q_1^2\sin^2\tgun}{q_1} \frac{\Delta\Pi_1}{\Delta\Pi_2} \\
\end{equation}

\section{Comparison between ADIPLS and GYRE \label{app_adipls_gyre}}

\begin{figure}[!htb]
   \begin{minipage}{0.48\textwidth}
     \centering
     \includegraphics[width=9cm]{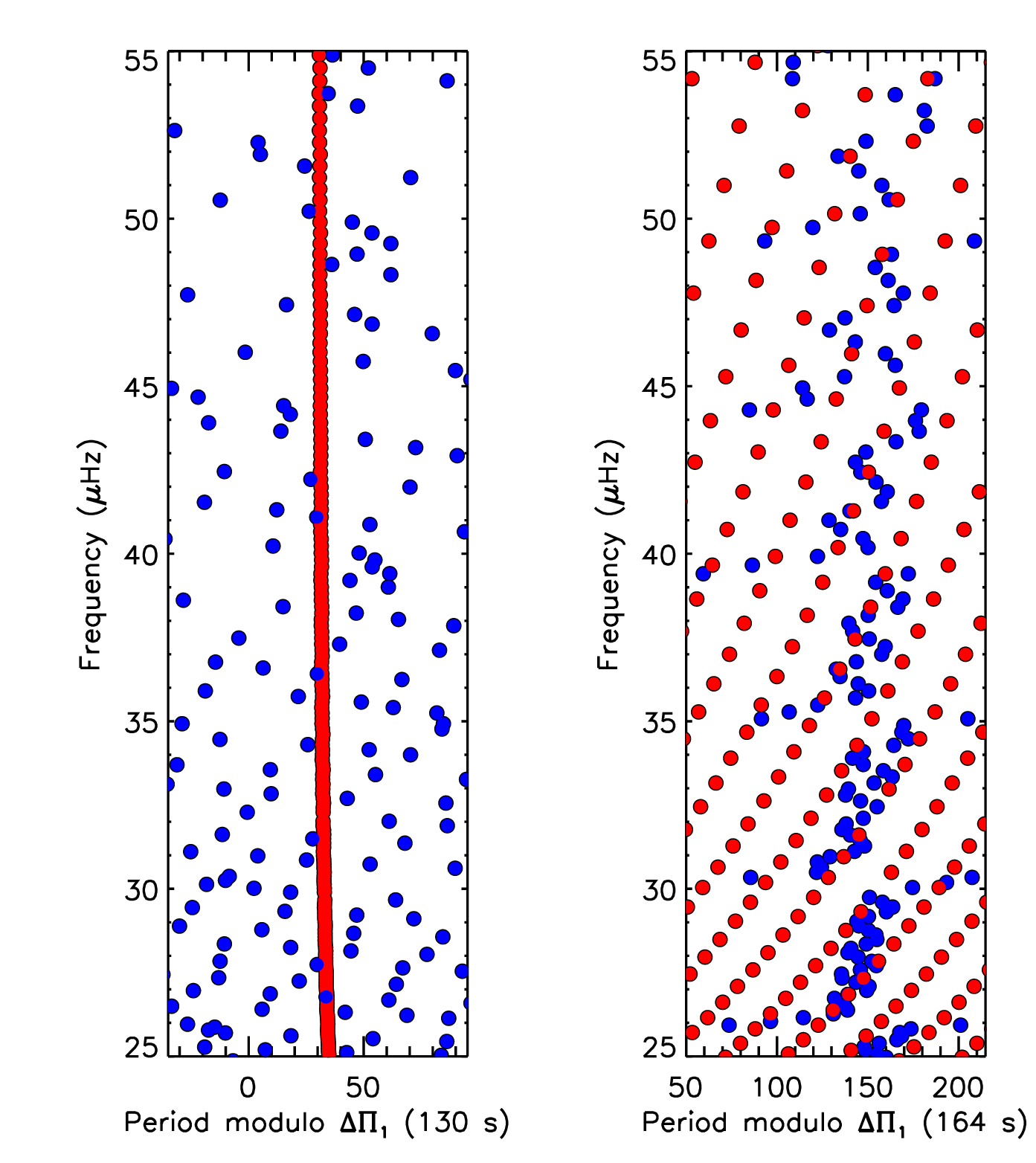}
     \caption{Period \'echelle diagrams of model 1 using oscillation modes computed with the code \gyre. These diagrams were obtained by folding the mode periods using alternately $\Delta\Pi_1$ (left panel) and $\Delta\Pi_2$ (right panel). Mode that have more the 50\% of the energy in the \gun-cavity (resp. \gdeux-cavity) are shown as filled red (resp. blue) circles. \label{fig_echelle_gyre}}
   \end{minipage}\hfill
   \begin {minipage}{0.48\textwidth}
     \centering
     \includegraphics[width=9cm]{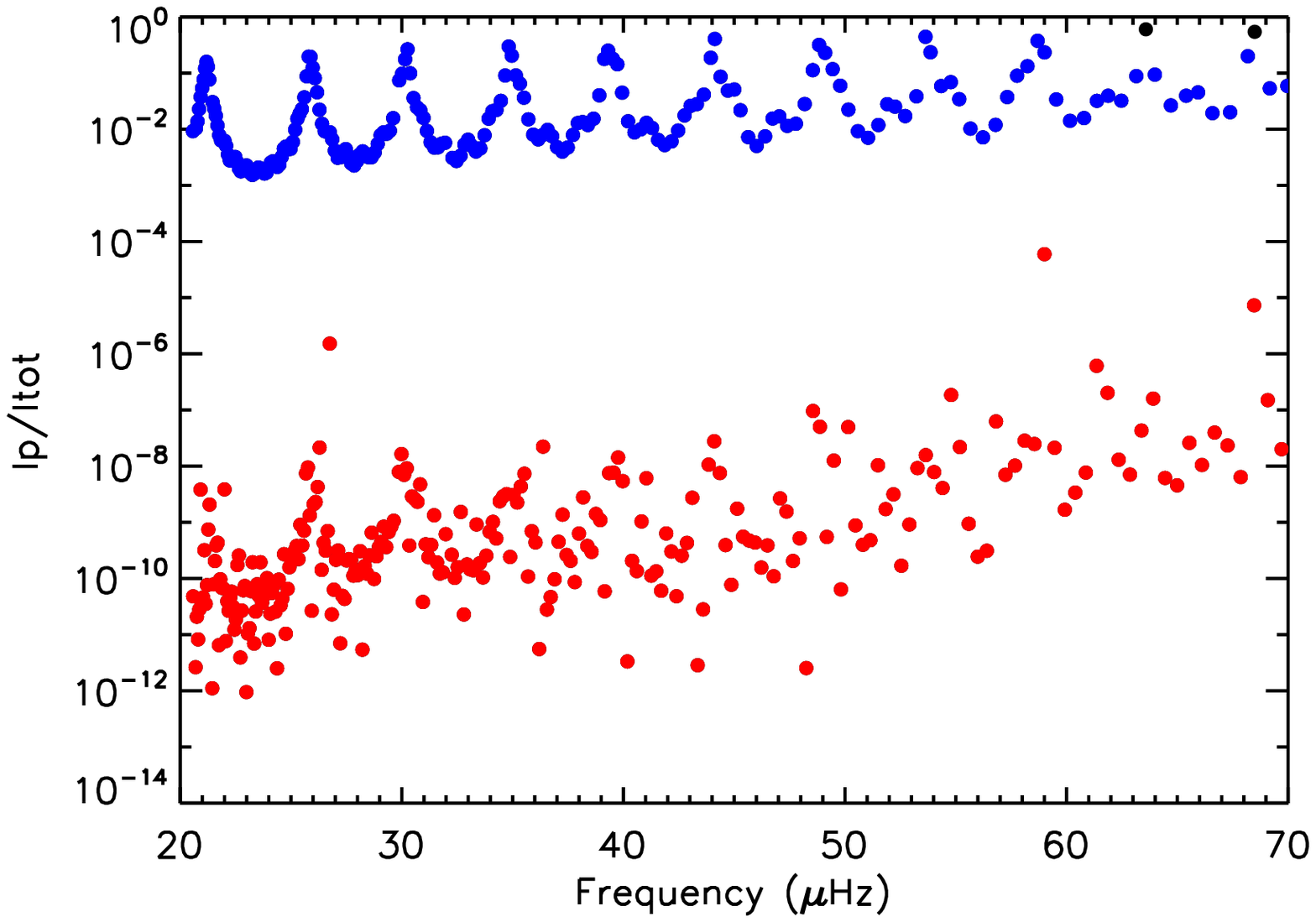}
     \caption{Fractional contribution of the acoustic cavity to the mode inertia for model 1, computed with \gyre. \label{fig_ratio_gyre}}
    \end{minipage}
\end{figure}

As we have mentioned in Sect. \ref{sect_1Dmodel}, numerically calculating the properties of oscillation modes during a He subflash is tricky. Oscillation codes have never been tested in this special case, and one must therefore treat the results with care. In order to validate the numerical frequencies and eigenfunctions of the modes obtained with the code \adipls, we also calculated the numerical frequencies and eigenfunctions of the modes using the code \gyre\ (\citealt{gyre}), which has been adapted to treat RGB stars. In particular, \gyre\ performs an interpolation of equilibrium quantities to ensure a certain number of points per mode wavelength in the propagation regions.

We show in Fig. \ref{fig_echelle_gyre} the period \'echelle diagrams calculated with the code \gyre\ for model 1. As is done in Sect. \ref{sect_numerical} with oscillation modes computed with \adipls, the period \'echelle diagrams were obtained by folding the mode periods with the asymptotic period spacing of the \gun-cavity $\Delta\Pi_1$ (left panel) and with the asymptotic period spacing of the \gdeux-cavity $\Delta\Pi_2$ (right panel). This figure is to be compared to the two top-left panels of Fig. \ref{fig_echelle_adipls}, which show period \'echelle diagrams calculated with \adipls\ for model 1. The \'echelle diagrams are very similar, showing a quite good overall agreement between the two codes concerning the pattern of the oscillation spectrum. A closer inspection shows slight mode-to-mode differences that will need to be looked into, but do not influence the conclusions of the present paper.

\begin{figure}
\begin{center}
\includegraphics[width=9cm]{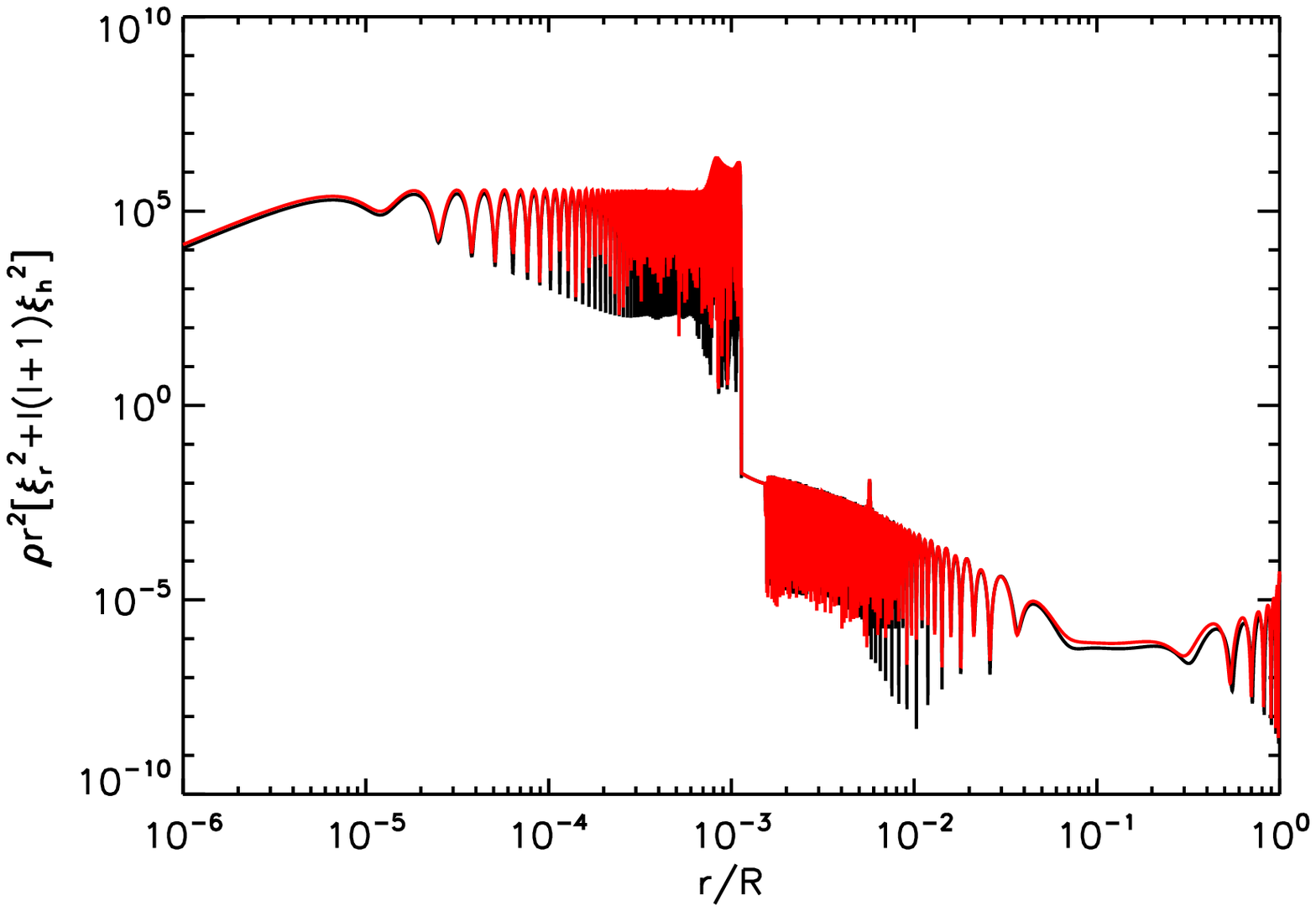}
\includegraphics[width=9cm]{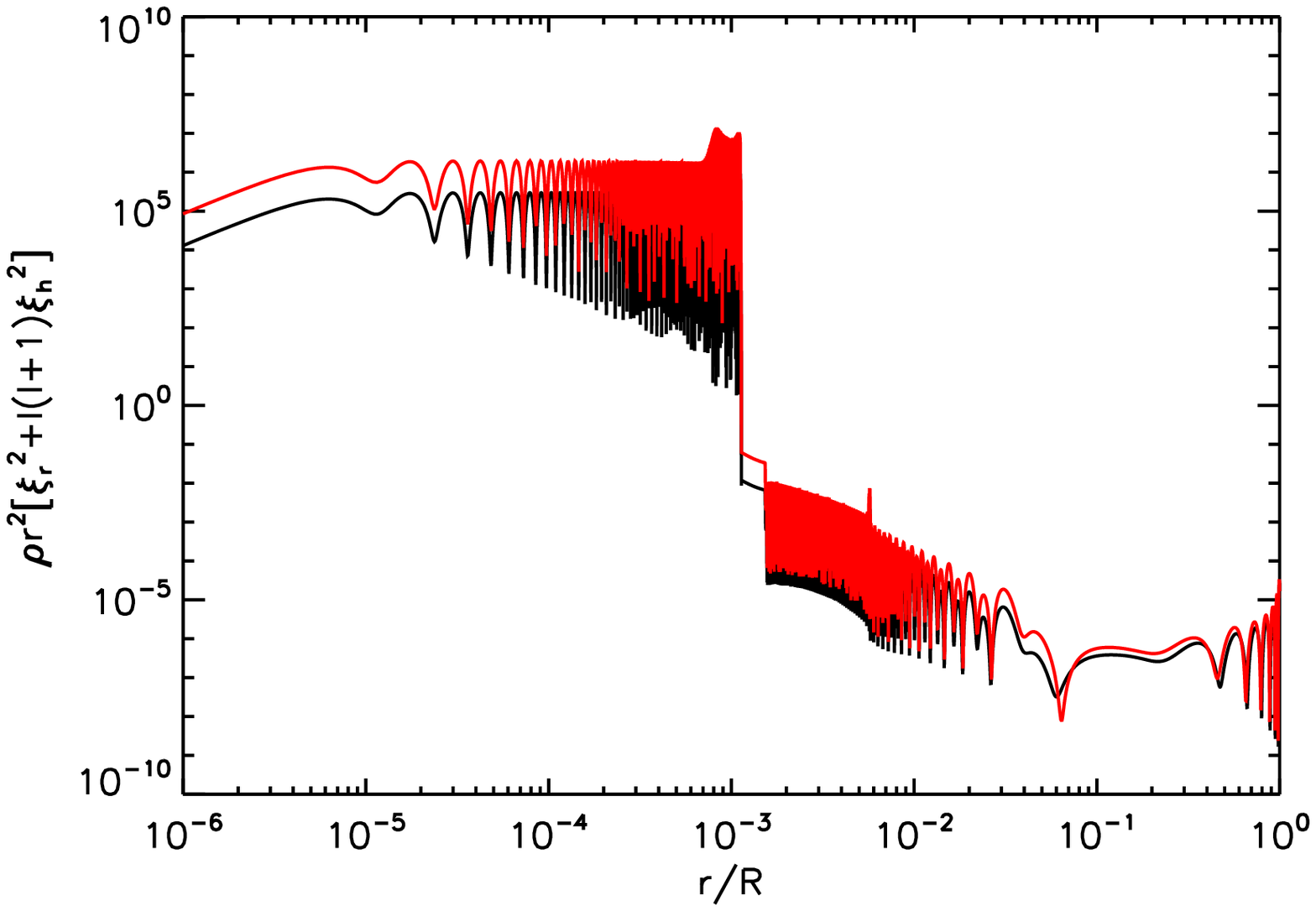}
\end{center}
\caption{Integrand of the inertia of two \gun-dominated dipolar mixed modes of model 1 computed with \adipls\ (red curves) and \gyre\ (black curves).
\label{fig_eigen_gyre}}
\end{figure}

A critical point in this study is the intensity of the coupling between the two g-mode cavities, which is reflected by the ratios of inertias of the modes in the different resonant cavities. In order to check the results of the code \adipls, we also calculated the ratio $I_{\rm p}/I$ for the modes of model 1 computed with \gyre. The results are shown in Fig. \ref{fig_ratio_gyre}. This figure is to be compared to the top left panel of Fig. \ref{fig_ratio_adipls}, which shows the same quantity for modes computed with \adipls. Once again, the overall agreement between the two codes is excellent. For both codes, the fractional contribution of the acoustic cavity to the total mode inertia is between $10^{-12}$ and $10^{-4}$ for \gun-dominated modes, and for \gdeux- and p-dominated modes, the output of the two codes are nearly identical. We do notice some mode-to-mode differences, whereby the contribution of the \gun-cavity to the mode inertia differs from one code to the other. We show in Fig. \ref{fig_eigen_gyre} the integrand of the inertia for two \gun-dominated mixed modes of model 1 computed with \adipls\ (in red) and \gyre\ (in black). For one of the modes, the two codes agree very well, while for the other, the contribution of the \gun-cavity to the mode inertia differs by a factor of 10 between the codes. These mode-to-mode differences will need to be investigated more thoroughly. However, the oscillation codes give an identical general pattern for the oscillations spectra and they yield similar order-of-magnitude values for the ratios of the mode inertia in the different propagating cavities. This shows that the results of the oscillation code \adipls\ can reasonably be trusted, to the level of precision required in this study.

%

\end{appendix}

\twocolumn

\end{document}